\documentclass[a4paper,11pt]{article}
\usepackage{jcappub}
\usepackage[T1]{fontenc}
\usepackage{graphicx}
\usepackage{amsmath}
\usepackage{amsmath}
\usepackage{xspace}
\usepackage{subfig}
\usepackage{float}
\usepackage{color}
\usepackage[utf8]{inputenc}
\usepackage{commath}
\usepackage{slashed}
\usepackage{fancyvrb}
\usepackage[utf8]{inputenc}
\usepackage[english]{babel}
\usepackage{amsfonts}
\usepackage{amssymb}
\usepackage{orcidlink}
\usepackage{color}
\usepackage{cancel}
\newcommand{\be}{\begin{equation}}
\newcommand{\ee}{\end{equation}}
\newcommand{\bea}{\begin{eqnarray}}
\newcommand{\eea}{\end{eqnarray}}

\catcode`@=12

\title{Anatomy of singlet-doublet dark matter relic: annihilation, co-annihilation, co-scattering, and freeze-in}

\author[a]{Partha Kumar Paul$^{\orcidlink{https://orcid.org/0000-0002-9107-5635}}$,}
\emailAdd{ph22resch11012@iith.ac.in}
\author[a]{Sujit Kumar Sahoo$^{\orcidlink{https://orcid.org/0000-0002-9014-933X}}$,}
\emailAdd{ph21resch11008@iith.ac.in}
\author[a]{Narendra Sahu$^{\orcidlink{https://orcid.org/0000-0002-9675-0484}}$,}
\emailAdd{nsahu@phy.iith.ac.in}
\affiliation[a]{Department of Physics, Indian Institute of Technology Hyderabad, Kandi, Telangana-502285, India.}

\abstract{The singlet-doublet vector-like fermion dark matter model has been extensively studied in the literature over the past decade. An important parameter in this model is the singlet-doublet mixing angle ($\sin\theta$). All the previous studies have primarily focused on annihilation and co-annihilation processes for obtaining the correct dark matter relic density, assuming that the singlet and doublet components decouple at the same epoch. In this work, we demonstrate that this assumption holds only for larger mixing angles with a dependency on the mass of the dark matter. However, it badly fails for the mixing angle $\sin\theta<0.05$. We present a systematic study of the parameter space of the singlet-doublet dark matter relic, incorporating annihilation, co-annihilation, and, for the first time, co-scattering processes. Additionally, non-thermal productions via the freeze-in and SuperWIMP mechanism are also explored. We found that due to the inclusion of co-scattering processes, the correct relic density parameter space is shifted towards the detection sensitivity range of the LHC and MATHUSLA via displaced vertex signatures.}


\begin{document}
	
	\maketitle
	\flushbottom
	
	\setcounter{footnote}{0}
	\renewcommand*{\thefootnote}{\arabic{footnote}}
	
	\section{Introduction}\label{intro}
	
	Dark matter (DM) is the mysterious form of matter that makes up about 26.8\% of the Universe's total energy density \cite{Planck:2018vyg,WMAP:2012nax}. Its existence on large scales has been firmly established through astrophysical evidence such as galaxy rotation curves, gravitational lensing, and the large-scale structure of the Universe. As the standard model (SM) of particle physics is unable to provide a suitable candidate for DM, particle physicists look beyond the SM (BSM) scenarios for DM. Weakly Interacting Massive Particle (WIMP) \cite{Lee:1977ua,Griest:1989wd,Kolb:1990vq} is one of the well-motivated BSM scenarios where the DM relic density is established due to the freeze-out of different processes in the early Universe.
	
	A simplest extension of the SM for a DM candidate is a vector-like singlet fermion: $\chi$ with an additional $\mathcal{Z}_2$ symmetry under which $\chi$ is odd while all other particles are even \cite{Bhattacharya:2018fus}. It has only a dimension five effective coupling with the SM Higgs boson {\it, i.e.} $(\bar{\chi}\chi H^\dagger H)/\Lambda$, where $\Lambda$ is the cutoff scale. A very large coupling is required to satisfy the correct relic density. However, the direct detection experiments exclude such large couplings. In the singlet fermionic DM model, only DM mass at the Higgs resonance ($M_{\rm DM}=M_H/2$) survives the direct detection constraints.
	On the other hand, if the SM is extended with a $\mathcal{Z}_2$-odd vector-like fermion doublet, then the neutral component of the doublet can be a viable DM candidate \cite{Bhattacharya:2018fus}. However, due to large gauge coupling, the annihilation cross-section is so large that the DM becomes under-abundant in the low mass region. The relic is satisfied at a very high DM mass, typically at the TeV scale. Moreover, the doublet DM model is ruled out by the current direct detection constraints \cite{LZ:2022lsv}.  This problem can be resolved if the singlet is combined with the vector-like doublet fermion, which are odd under a $\mathcal{Z}_2$ symmetry. In this case, the neutral component of the doublet mixes with the singlet to give a singlet-doublet DM (SDDM). By choosing the SD mixing appropriately, one can get the correct relic density in a large range of parameter space, which also satisfies the direct detection constraints. The SDDM has been studied extensively in the literature \cite{Cynolter:2015sua,Bhattacharya:2015qpa,Bhattacharya:2018fus,Bhattacharya:2017sml,Bhattacharya:2018cgx,Bhattacharya:2016rqj,Dutta:2020xwn,Borah:2021khc,Borah:2021rbx,Borah:2022zim,Borah:2023dhk,Paul:2024iie,Mahbubani:2005pt,DEramo:2007anh,Cohen:2011ec,Freitas:2015hsa,Calibbi:2015nha,Cheung:2013dua,Banerjee:2016hsk,DuttaBanik:2018emv,Horiuchi:2016tqw,Restrepo:2015ura,Abe:2017glm,Konar:2020wvl,Konar:2020vuu,Calibbi:2018fqf,Ghosh:2021wrk,Das:2023owa,Bhattacharya:2021ltd,Enberg:2007rp}. An interesting feature of this model is that it introduces co-annihilation processes along with the annihilation among the dark sector particles to give rise the correct relic density of DM. The two most important parameters in the SDDM model are the SD mixing angle ($\sin\theta$) and the mass splitting ($\Delta{M}$) between the DM (dominantly singlet) and the NLSP (dominantly the doublet). For a larger $\sin\theta$ and $\Delta M$, the annihilation process dominantly decides the relic density of the DM. On the other hand, in the small mass-splitting regime, the co-annihilation process becomes dominant.	In previous studies of the SDDM \cite{Bhattacharya:2015qpa,Bhattacharya:2018fus,Bhattacharya:2017sml,Bhattacharya:2018cgx,Bhattacharya:2016rqj,Dutta:2020xwn,Borah:2021khc,Borah:2021rbx,Borah:2022zim,Borah:2023dhk,Paul:2024iie}, it was generally assumed that the dark sector particles decouple from the thermal bath at the same epoch. However, this assumption may not hold true for small mixing angles; typically, for $\sin\theta\lesssim0.05$, the DM particles are not in chemical equilibrium with the thermal bath. In particular, the singlet component of the dark sector decouples from the thermal bath at an earlier time, while the doublet component remains in equilibrium for a longer duration due to its interactions with the SM gauge bosons. 	This implies that for $\sin\theta\lesssim0.05$, the relic density parameter space estimated in earlier studies is not valid, while they remain valid for $\sin\theta\gtrsim0.05$.
	
	In this paper, we revisit the SDDM relic density in the small mixing limits, typically $\mathcal{O}(10^{-7})\lesssim\sin\theta\lesssim0.05$ \footnote{If the SD mixing is large, typically $\mathcal{O}(1)$ and $\Delta{M}$ < 1 GeV, then the non-perturbative effects would be important in the relic density calculation, see \textit{e.g.} \cite{Hisano:2002fk,Oncala:2021tkz,Garny:2021qsr}. Here, for simplicity, we took $\Delta{M}\ge 1$ GeV.}. In this range of SD mixing, the singlet can be brought to equilibrium quickly. However, the DM (dominantly singlet) decouples early with a larger abundance.
	We demonstrate that by incorporating conversion-driven processes \cite{DAgnolo:2017dbv, Garny:2017rxs}, the DM relic density can be brought into agreement with the observed value. In particular, the kinetic equilibrium between DM and the thermal bath allows for upscattering processes, where DM particles can be scattered into a higher-mass state within the dark sector, such as the doublet, thereby reducing the singlet's abundance. Recent studies have explored the conversion-driven processes in various BSM contexts; see, for instance, \cite{Alguero:2022inz,Heeck:2022rep,Heisig:2024mwr,DiazSaez:2024nrq,  Zhang:2024sox}.  
	
	We further explore the DM parameter space by reducing the SD mixing angle. We saw that for $\sin\theta<\mathcal{O}(10^{-7})$, the singlet never reaches equilibrium, and hence the DM relic density can not be realized via the freeze-out mechanism. In such a small mixing angle regime, the DM can be produced non-thermally from the decay of the doublet partner. In Section \ref{sec:freezein}, we show that for DM mass ranging from 1 to 1000 GeV and $\Delta M\ge 1$ GeV, the required $\sin\theta$ for correct relic density varies from $\mathcal{O}(10^{-11})$ to $\mathcal{O}(10^{-13})$. On the hand, for SD mixing in the range, $\mathcal{O}(10^{-11})<\sin\theta<\mathcal{O}(10^{-7})$, the DM is over produced. For $\sin\theta<\mathcal{O}(10^{-14})$, DM relic can be achieved via the SuperWIMP mechanism \cite{Garny:2018ali,Junius:2019dci,Borah:2021rbx} from the out-of-equilibrium decay of the doublet components.
	
	In Section \ref{sec:dispacedvertex}, we also demonstrate that the charged partner of the SDDM can yield distinctive signatures at colliders. The doublet fermion can be produced copiously due to their gauge interaction. Once produced, they can decay to DM and leptons via the singlet doublet mixing, which can give
	displaced vertex signatures in the  LHC \cite{Cepeda:2019klc} and, in future, MATHUSLA \cite{MATHUSLA:2019qpy}.
	
	The paper is organized as follows. We discuss the SDDM model in section \ref{sec:model}. In section \ref{sec:annicoanni}, we study the correct relic parameter space considering annihilation and co-annihilation only. We study the effects of conversion-driven processes in section \ref{sec:co-sactter}. The freeze-in production of DM is discussed in section \ref{sec:freezein}. The production of DM relic via SuperWIMP mechanism is discussed in section \ref{sec:sumerwimp}. The constraint on the DM parameter space using the null detection of the DM at direct search experiments is discussed in section \ref{sec:DD}. Section \ref{sec:dispacedvertex} discusses the displaced vertex signatures. We finally conclude in section \ref{sec:conclu}.

	\section{The singlet-doublet dark matter model}\label{sec:model}
	
	In this case, the SM is extended with a fermion doublet $\Psi=(\psi^0~~\psi^{-})^T$ and a singlet fermion $\chi$, which are stabilized by a discrete symmetry $\mathcal{Z}_2$, under which they are odd, and all other SM particles are charged even \cite{Bhattacharya:2018fus}.
	
	The relevant Lagrangian of the model guided by imposed symmetry is given by
	\begin{eqnarray}
		\mathcal{L} &\supset& i \overline{\Psi} \gamma^\mu D_\mu \Psi + i \overline{\chi} \gamma^\mu \partial_\mu \chi - M_{\Psi} \overline{\Psi} \Psi -M_{\chi} \overline{\chi} \chi -
		y\overline{\Psi} \Tilde{H} \chi+ h.c.,
		\label{eq:lag}
	\end{eqnarray}
	where $D_\mu=\partial_\mu-g\frac{\tau_i}{2} W^i_\mu-g^\prime\frac{Y}{2} B_\mu$ and $H$ is the SM Higgs doublet.
	
	The scalar potential of the model is given by
	\begin{eqnarray}
		V_{\rm scalar} &=&  -\mu^2_H (H^\dagger H) + \lambda_H (H^\dagger H)^{2}.
		\label{scalarL}
	\end{eqnarray}
	After the electro-weak symmetry breaking, the quantum fluctuation around the minima is given as
	\begin{eqnarray}
		H=\begin{pmatrix}
			0\\
			\frac{v+h}{\sqrt{2}}
		\end{pmatrix}.
	\end{eqnarray}
	
	Once Higgs gets a vacuum expectation value, $v$, it induces mixing between singlet and neutral component of doublet fermion through $\overline{\Psi} \Tilde{H} \chi $-coupling. Denoting the mass eigenstates as $\chi_0$ and $\chi_1$ with the mixing angle $\sin\theta$, the transformation from the flavor states to physical states can be written as,
	
	\begin{eqnarray}
		\left(\begin{matrix}
			\psi^0 \\ \chi
		\end{matrix}\right)
		=\left(\begin{matrix}
			\cos\theta & -\sin\theta \\
			\sin\theta & \cos\theta
		\end{matrix}\right)
		\left(\begin{matrix}
			\chi_0 \\  \chi_1
		\end{matrix}\right),
	\end{eqnarray}
	where the mixing angle is given by
	\begin{eqnarray}\label{eq:mixang}
		\tan{2\theta}=  \frac{\sqrt2 y v}{M_\Psi-M_{\chi}} .
	\end{eqnarray}
	
	The mass eigenvalues of the physical states are given as
	
	\begin{eqnarray}
		M_{\chi_0}&=&M_{\Psi} \cos ^2 \theta + \frac{yv}{\sqrt{2}}\sin 2\theta + M_\chi \sin ^2 \theta,~~ \nonumber\\
		M_{\chi_1}&=&M_{\Psi} \sin ^2 \theta - \frac{yv}{\sqrt{2}}\sin 2\theta + M_\chi \cos ^2 \theta\equiv M_{\rm DM}
	\end{eqnarray}
	with a mass-splitting $\Delta M=M_{\chi_0}-M_{\chi_1}$, where, $\chi_1$ is dominantly the singlet fermion and $\chi_0$ is the $\psi^0$.
	From Eq. (\ref{eq:lag}), the interaction among the mass eigenstates ($\chi_0$, $\psi^\pm$ and $\chi_1$), can be expressed as,
	\begin{eqnarray}
		\mathcal{L}_{int} &=& \left(\frac{e_0}{2 \sin\theta_W \cos\theta_W}\right) \left[\sin^2\theta \overline{\chi_1}\gamma^{\mu}Z_{\mu}\chi_1+\cos^2\theta \overline{\chi_0}\gamma^{\mu}Z_{\mu}\chi_0 \right.\nonumber \\
		&{}&\left.+\sin\theta \cos\theta(\overline{\chi_1}\gamma^{\mu}Z_{\mu}\chi_0+\overline{\chi_0}\gamma^{\mu}Z_{\mu}\chi_1)\right]   \nonumber \\
		&{}&+\frac{e_0}{\sqrt2\sin\theta_W}\sin\theta \overline{\chi_1}\gamma^\mu W_\mu^+ \psi^- +\frac{e_0}{\sqrt2 \sin\theta_W} \cos\theta \overline{\chi_0}\gamma^\mu W_\mu^+ \psi^-  \nonumber \\
		&{}&  +\frac{e_0}{\sqrt2 \sin\theta_W} \sin\theta{\psi^+}\gamma^\mu W_\mu^- \chi_1 + \frac{e_0}{\sqrt2\sin\theta_W}\cos\theta {\chi^+}\gamma^\mu W_\mu^- \chi_0  \nonumber \\
		&{}&- \left(\frac{e_0}{2 \sin\theta_W\cos\theta_W}\right) \cos2\theta_W {\psi^+}\gamma^{\mu}Z_{\mu}\psi^- - e_0 {\psi^+}\gamma^{\mu}A_{\mu}\psi^- \nonumber \\
		&{}& -\frac{y}{\sqrt2}h\left[\sin2\theta(\overline{\chi_1}\chi_1-\overline{\chi_0}\chi_0)+\cos2\theta(\overline{\chi_1}\chi_0+\overline{\chi_0}\chi_1)\right],
	\end{eqnarray}
	
	where $e_0=0.313$ is the electromagnetic coupling constant, and $\theta_W$ is the Weinberg angle. Note that in this model, there are three parameters: $M_{\rm DM},\sin\theta, \Delta M$, which decide the relic. In the rest of the draft, we will find the appropriate parameter space for which the correct relic density will be obtained.
	
	\section{Dark matter phenomenology}\label{sec:DMpheno}
	
In this section, we systematically explore the relic density parameter space for SDDM, considering annihilation, co-annihilation, co-scattering, and freeze-in, depending on the $\sin\theta$ and $\Delta{M}$. We divide the parameter space into four regimes: (i) thermal DM relic considering annihilation and co-annihilation only, (ii) thermal DM relic via conversion-driven processes, (iii) dark matter production via freeze-in mechanism, and (iv) dark matter production via SuperWIMP mechanism.
	
	\subsection{Thermal production of dark matter}\label{sec:ann-coann}
	
	Here, we assume that the DM is in thermal equilibrium with the SM bath particles. This can be realized by considering a large SD mixing. The SD mixing is chosen in such a way that all the dark sector particles decouple at the same epoch. The relic density of DM in this scenario can be achieved by solving the Boltzmann equation (BE) \cite{Griest:1990kh},
	\begin{eqnarray}{\label{eq:BE1}}
		\frac{dn}{dt} + 3 \mathcal{H} n = - \langle \sigma v \rangle_{\rm eff} (n^{2} - (n^{\rm eq})^{2})
	\end{eqnarray}
	where $n=\sum_i n_i $ represents the total number density of all dark sector particles,  $n^{\rm eq}$ is the corresponding equilibrium number density, and the Hubble parameter is expressed as $\mathcal{H}=\sqrt{\frac{4\pi^3g_{*\rho}}{45M^2_{\rm Pl}}}T^2$ with the Planck mass, $M_{\rm Pl}=1.22\times10^{19}$ GeV, and $g_{*\rho}$ is the total relativistic contribution to the energy density.  $\langle \sigma v \rangle_{\rm eff}$ represents the effective cross-section which takes into account all number changing processes for DM freeze-out and is given by
	\begin{eqnarray}
		\langle \sigma v \rangle_{\rm eff} &=& \frac{g^2_1}{g^2_{\rm eff}}\langle\sigma v\rangle_{\chi_{_1}\chi_{_1}} + \frac{g_0 g_1}{g^2_{\rm eff}}\langle\sigma v\rangle_{\chi_{_0}\chi_{_1}} (1+\Delta_{\chi_{_0}})^{{3}/{2}}\exp(-x\Delta_{\chi_{_0}}) \nonumber\\&{}&+\frac{g_2 g_1}{g^2_{\rm eff}}\langle\sigma v\rangle_{\chi_{_1}\psi^{-}} (1+\Delta_{\psi^{-}})^{{3}/{2}}\exp(-x\Delta_{\psi^{-}})\nonumber\\&{}&+\frac{g^2_1}{g^2_{\rm eff}}\langle\sigma v\rangle_{\chi_{_0}\chi_{_0}} (1+\Delta_{\chi_{_0}})^{3}\exp(-2x\Delta_{\chi_{_0}})+\frac{g^2_2}{g^2_{\rm eff}}\langle\sigma v\rangle_{\psi^{+}\psi^{-}} (1+\Delta_{\psi^{-}})^{3}\exp(-2x\Delta_{\psi^{-}}) \nonumber\\&{}&+\frac{g_0 g_{2}}{g^2_{\rm eff}}\langle\sigma v\rangle_{\chi_{_0} \psi^{-}} (1+\Delta_{\chi_{_0}})^{3/2}(1+\Delta_{\psi^{-}})^{3/2}\exp(-x(\Delta_{\chi_{_0}}+\Delta_{\psi^{-}}))
		\label{eq:effcrs}
	\end{eqnarray}
	where $g_0, g_1$ and $ g_2$ represent the internal d.o.f. of $\chi_0,\chi_1$ and $\psi^{-}$ respectively and $\Delta_i$ stands for the ratio $(M_i-M_{\chi_1})/M_{\chi_{1}}$ with $M_i$ denoting the mass of $\chi_0$ and $\psi^{-}$. Here $g_{\rm eff}$ is the effective degree of freedom which is given by
	\begin{eqnarray}
		g_{\rm eff}= g_1 +g_0 (1+\Delta_{\chi_{_0}})^{3/2}\exp(-x\Delta_{\chi_{_0}})+ g_2 (1+\Delta_{\psi^{-}})^{3/2}\exp(-x\Delta_{\psi^{-}})
	\end{eqnarray}
	and $x$ is the dimensionless parameter defined as $x=M_{\rm \chi_1}/T$.
	The relic density of DM $\chi_1$ can then be obtained as
	\begin{eqnarray}
		\Omega_{\chi_1} h^2\equiv\Omega_{\rm DM} h^2 = \frac{1.09 \times 10^9 {\rm GeV}^{-1}}{\sqrt{g_* }M_{\rm Pl}} \left[\int_{x_f}^\infty dx~\frac{\langle \sigma v \rangle_{\rm eff}}{x^2}\right]^{-1}.
	\end{eqnarray}
	Here $x_f =M_{\chi_1}/T_{f}$, and $T_f$ denotes the freeze-out temperature of $\chi_1$.
	
	Upon analyzing Eq. (\ref{eq:effcrs}), the first three terms correspond to the annihilation and co-annihilation of $\chi_1$, while the last three terms describe the annihilation and co-annihilation processes involving $\chi_0$ and $\psi^-$ (The respective processes are listed in Appendices \ref{app:ann}, \ref{app:coann} and \ref{app:ann_doublet}.). The latter terms are associated with gauge-mediated interactions and are independent of the mixing angle: $\sin\theta$. Consequently, these processes dominate over the first three terms in the regime of smaller values of $\sin\theta$. Therefore, it is essential to verify whether the DM remains in chemical equilibrium with the $\mathcal{Z}_2$-odd doublet fermion in this low $\sin\theta$ region. Previous studies \cite{Bhattacharya:2018fus} have implicitly assumed that the dark sector particles ($\chi_1,\chi_0,\psi^-$) decouple at the same epoch for $\sin\theta\sim\mathcal{O}(10^{-2})$. In this work, we first verify the validity of this assumption and find the lowest value of $\sin\theta$ for which this assumption is valid. In other words, we find for what values of $\sin\theta$ the dark sector particles chemically decouple at the same epoch. We then show that even for small mixing angles, when the dark sector particles chemically decouple at different epochs, the correct relic abundance can still be achieved through processes known as co-scattering. In this mechanism, the chemically decoupled DM continues to scatter with the SM bath, producing heavier dark-sector states. In our scenario, the singlet DM can scatter off SM particles to produce doublet states. A detailed list of such processes is provided in the Appendix \ref{app:co-scattering}. However, such processes are not included in Eq. (\ref{eq:BE1}).

In order to incorporate such co-scattering processes, we define two dark sectors: (a) sector 1, which contains $\chi_1$, and (b) sector 2, which contains $\chi_0,\psi^-$, whereas all the SM particles are assigned as sector 0. The singlet fermion, being the lightest among the dark sector particles, serves as the DM candidate. By defining the comoving number density of sector 1 and sector 2 particles as $Y_1\left(=\frac{n_{\chi_1}}{s}\right)$ and $Y_2\left(=\frac{n_{\chi_0}+n_{\psi^\pm}}{s}\right)$ \footnote{Processes like $\psi^- SM \rightarrow\chi_0 SM$ will keep the neutral and charged components of the doublet in equilibrium always which makes the abundances of sector 2 particles to be the same.}, respectively, the general coupled BEs for their evolution are given as \begin{eqnarray}
		\frac{dY_1}{dT} &=&   \frac{1}{3\mathcal{H}}\frac{ds}{dT} \left[    \langle \sigma_{1100} v \rangle ( Y_1^2 - {Y_1^{\rm eq}}^2) +    \langle \sigma_{1122} v \rangle \left( Y_1^2 - Y_2^2  \frac{{Y_1^{\rm eq}}^2}{{Y_2^{\rm eq}}^2}\right)  + \langle \sigma_{1200} v \rangle ( Y_1 Y_2 - Y_1^{\rm eq}Y_2^{\rm eq})\right. \nonumber\\
		&&+\left.  \langle \sigma_{1222} v \rangle \left( Y_1 Y_2 - Y_2^2   \frac{Y_1^{\rm eq}}{Y_2^{\rm eq}} \right) -\langle \sigma_{1211} v \rangle \left( Y_1 Y_2 - Y_1^2   \frac{Y_2^{\rm eq}}{Y_1^{\rm eq}} \right)
		-\frac{ \Gamma_{2\rightarrow 1}}{s}\left( Y_2 -Y_1 \frac{Y_2^{\rm eq}}{Y_1^{\rm eq}}  \right)        \right] ,\nonumber\\
		\label{eq:Y1}
	\end{eqnarray}
	\begin{eqnarray}
		\frac{dY_2}{dT} &=&   \frac{1}{3\mathcal{H}}\frac{ds}{dT}\left[    \langle \sigma_{2200} v \rangle ( Y_2^2 - {Y_2^{\rm eq}}^2) -    \langle \sigma_{1122} v \rangle \left( Y_1^2 - Y_2^2  \frac{{Y_1^{\rm eq}}^2}{{Y_2^{\rm eq}}^2}\right) +  \langle \sigma_{1200} v \rangle ( Y_1 Y_2 - Y_1^{\rm eq}Y_2^{\rm eq}) \right. \nonumber \\
		&&- \left. \langle \sigma_{1222} v \rangle \left( Y_1 Y_2 - Y_2^2   \frac{Y_1^{\rm eq}}{Y_2^{\rm eq}} \right)
		+\langle \sigma_{1211} v \rangle \left( Y_1 Y_2 - Y_1^2   \frac{Y_2^{\rm eq}}{Y_1^{\rm eq}} \right)  + \frac{ \Gamma_{2\rightarrow 1}}{s}\left( Y_2 -Y_1 \frac{Y_2^{\rm eq}}{Y_1^{\rm eq}}  \right)        \right],\nonumber\\
		\label{eq:Y2}
	\end{eqnarray}
	where $Y_i^{\rm eq}\left(=\frac{n_i^{\rm eq}}{s}\right)$ are the equilibrium abundances, $\mathcal{H}$ is the  Hubble parameter, the entropy density, $s=\frac{2\pi^2}{45}g_{*s}T^3$, $\langle \sigma_{\alpha\beta\gamma\delta} v\rangle$ are the thermally averaged cross-sections for processes involving annihilation of particles of sectors $\alpha\beta\rightarrow \gamma\delta$, which is given by \cite{Gondolo:1990dk}
    \begin{eqnarray}
      \langle \sigma_{\alpha\beta\gamma\delta} v\rangle=\frac{T}{8m_{\alpha}^2m_{\beta}^2K_{2}(\frac{m_\alpha}{T})K_{2}(\frac{m_\beta}{T})}\int_{(m_\alpha+m_\beta)^2}^\infty\sigma_{\alpha\beta\rightarrow\gamma\delta}(s)\big(s-(m_\alpha+m_\beta)^2\big)\sqrt{s}K_1\bigg(\frac{\sqrt{s}}{T}\bigg)ds,\nonumber\\ 
    \end{eqnarray}
and $\Gamma_{2\rightarrow 1}$ is the conversion term, which includes both the interaction rate of the co-scattering process and the decay defined as   
\begin{eqnarray}\label{eq:gamma21}
\Gamma_{2\rightarrow1}=\Gamma_{\Psi\rightarrow\chi,\rm SM}\frac{K_1(M_{\Psi}/T)}{K_2(M_{\Psi}/T)}+\langle \sigma_{2010} v \rangle n^{\rm eq}_{\rm SM},
    \end{eqnarray}
where $\Gamma_{\Psi\rightarrow\chi,\rm SM}$ includes two body decays like $\Gamma_{\chi_0\rightarrow\chi_1 H},\Gamma_{\chi_0\rightarrow\chi_1 Z},\Gamma_{\psi^\pm\rightarrow\chi_1 W^{\pm}}$, and three body decay of the doublets as $\Gamma_{\psi^\pm\rightarrow\chi_1 l^\pm\nu_l}$\footnote{The decay mode $\Gamma_{\psi^\pm\rightarrow\chi_1 \pi^\pm}$ is very suppressed. This is because of $\Delta M\ge1$ GeV.}. The decay rates are given in Appendix \ref{app:decayrate}. $\langle \sigma_{2010} v \rangle$ is the thermally averaged cross-section of the co-scattering processes with $n_{\rm SM}^{\rm eq}=0.238\times s$ \cite{Alguero:2023zol}.
\subsubsection{Thermal DM relic in the large mixing regime via annihilation and co-annihilation}\label{sec:annicoanni}

We are now well equipped to calculate the thermal relic in the both small and large $\sin\theta$ regime. As discussed before for $\sin\theta\sim\mathcal{O}(10^{-2})$, the dark sector particles decouple chemically at the same epoch. To validate this assumption of chemical equilibrium among the dark sector particles we define a new parameter $\rho_1$ as
	\begin{equation}    \rho_1=\frac{\Omega_{1s}h^2}{\Omega_{2s}h^2(\rm no ~coscattering)},
		\label{eq:rho1}
	\end{equation}
	where $\Omega_{1s}h^2$ is the relic density obtained by solving Eq. (\ref{eq:BE1}), and $\Omega_{2s}h^2$(no co-scattering) is calculated by solving Eqs. (\ref{eq:Y1}) and (\ref{eq:Y2}) excluding co-scattering processes\footnote{We use {\tt micrOMEGAs 6.1.15}\cite{Alguero:2023zol,Alguero:2022inz} to solve the Boltzmann equations, where {\tt darkOmega} function is used to estimate $\Omega_{1s}h^2$ and {\tt darkOmegaN} is used to estimate $\Omega_{2s}h^2$(no co-scattering) while excluding the co-scattering processes. This can be achieved by using the flag {\tt ExcludedForNDM="2010"}, while using {\tt darkOmegaN} function.}. In particular we have switched off $\langle\sigma_{2010}v\rangle n^{\rm eq}_{\rm SM}$ in Eq. (\ref{eq:gamma21}). In Eq. (\ref{eq:rho1}), $\rho_1=1$ implies that all the dark sector particles decouple at the same epoch. On the other hand,	$\rho_1<1$ implies that the dark sector particles are decoupling at different epochs. When $\sin\theta$ is small, the singlet decouples earlier than the doublet and produces a relatively larger relic in absence of the co-scattering processes. This implies that $\Omega_{2s}h^2({\rm no~coscatttering})>\Omega_{1s}h^2$ Therefore, one expects $\rho_1\le1$.
	
	\begin{figure}[h]
		\centering     \includegraphics[width=7.5cm,height=7.5cm]{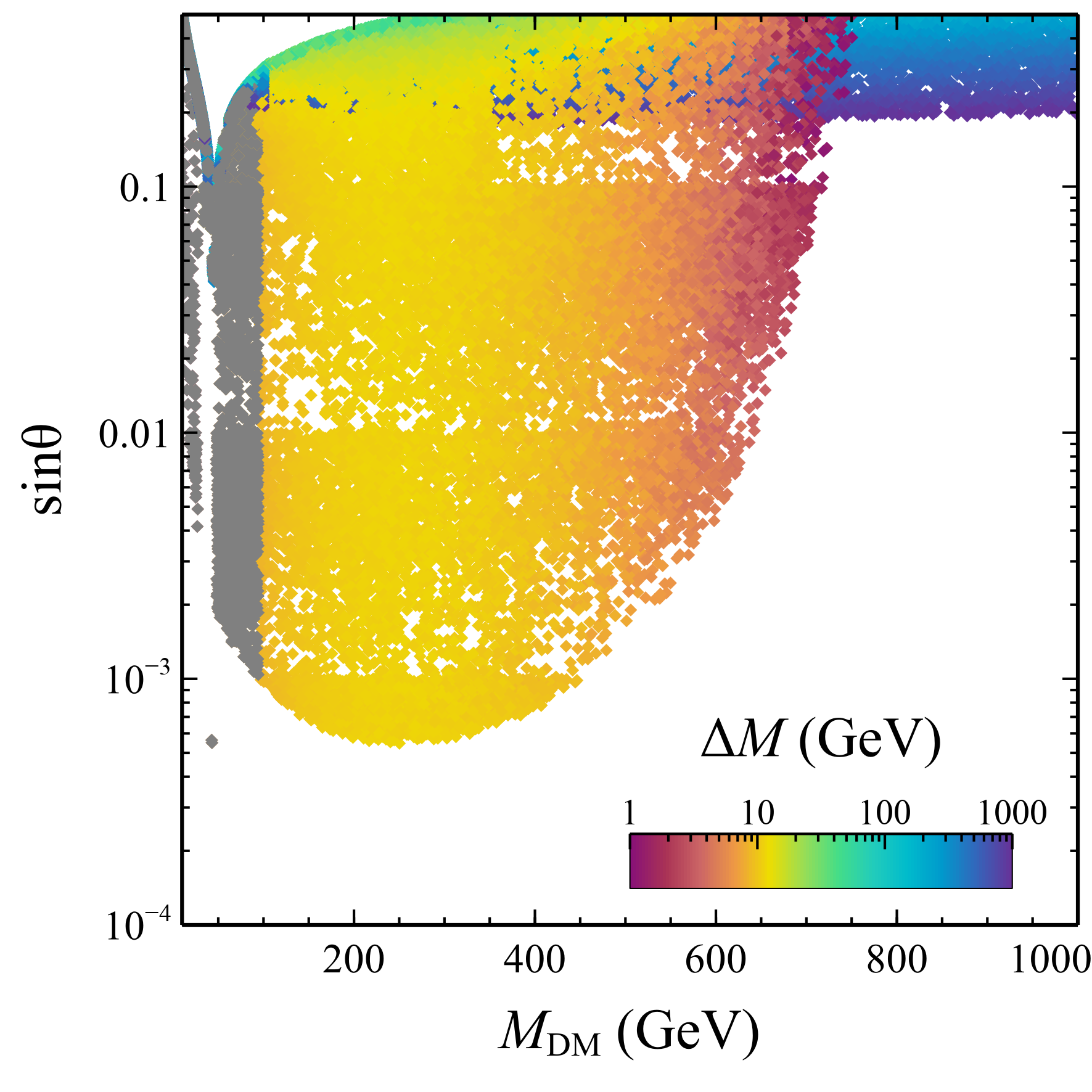}        \includegraphics[width=7.5cm,height=7.5cm]{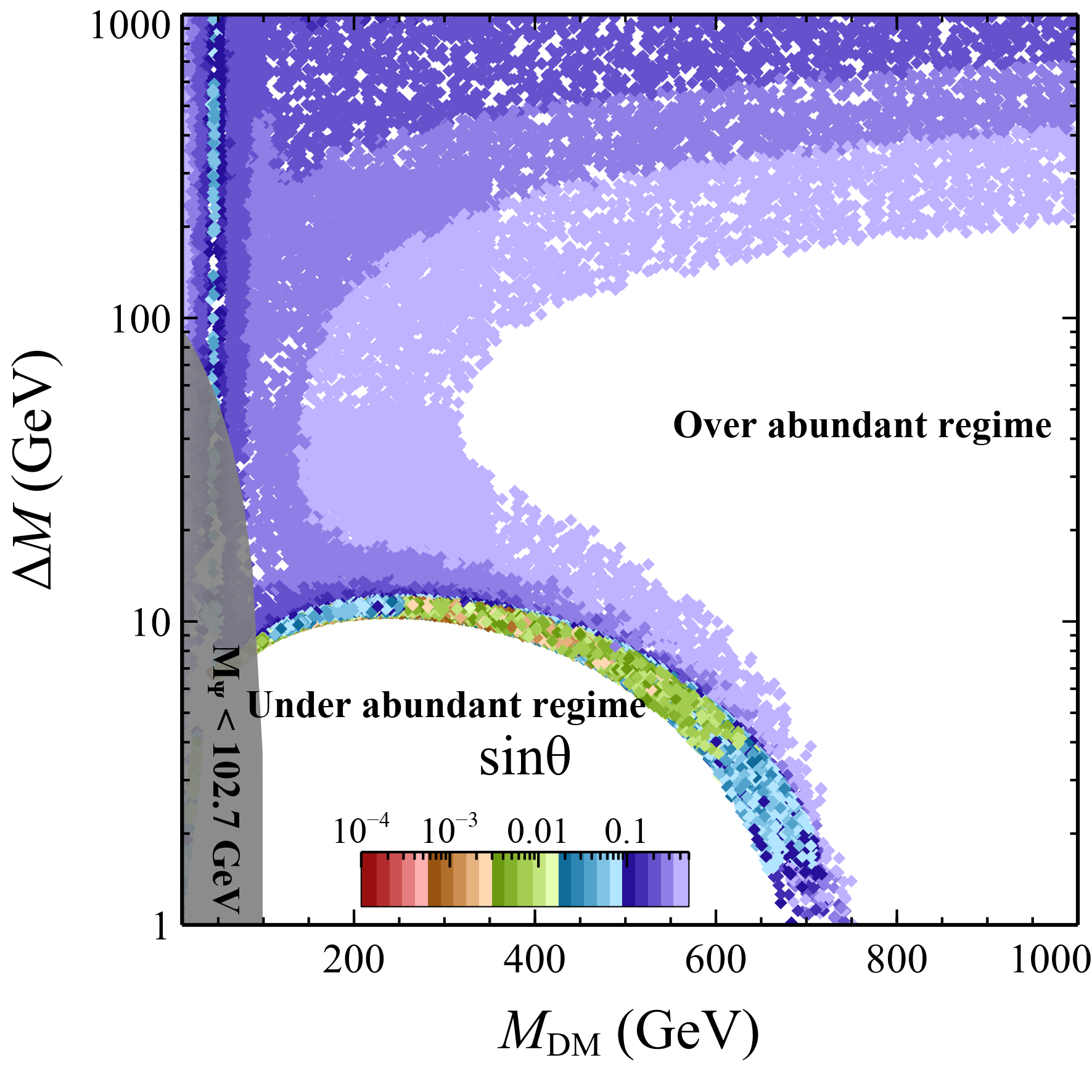}
		\caption{Allowed parameter space of DM for which DM is in equilibrium with SM bath and the $\mathcal{Z}_2$-odd doublet. All the points in the plots (\textit{left} and \textit{right}) satisfy the required relic density. The grey-shaded region in both the plots shows the LEP bound \cite{DELPHI:2003uqw} on the heavy fermion doublet.}
		\label{fig:case1_anni}
	\end{figure}
	We have presented the parameter space in Fig. \ref{fig:case1_anni}. In this plot, all the points satisfy the correct relic density of DM, and we have used $0.99<\rho_1\le1$ to assess the validity and domain of applicability of the Eq. (\ref{eq:BE1}). In Fig. \ref{fig:case1_anni} ({\it left}), we have shown all the points in the plane of $\sin\theta$ and $M_{\rm DM}$, and in the {\it right} plot, we projected the same points in the plane of $\Delta{M}$ and $M_{\rm DM}$. DM mass is varied in the range of $1-1000$ GeV, $\Delta{M}$ is varied in the range of $1-1000$ GeV, and $\sin\theta$ is varied in the range of ($0.5-10^{-4}$). From Fig. \ref{fig:case1_anni} ({\it left}), we observe that the value of $\sin\theta$ can be as small as $\sim5\times10^{-4}$ for $M_{\rm DM}\sim 250$ GeV. As shown in Fig. \ref{fig:case1_anni} ({\it right}), we exactly reproduced all the points that satisfy correct relic density in the plane of $\Delta{M}$ vs $M_{\rm DM}$ as given in \cite{Bhattacharya:2018fus}. The pattern shown in the right plot arise from the interplay between annihilation and co-annihilation processes. In the region with small mass splitting ($\Delta{\rm M}$ $\lesssim15$ GeV), the relic density satisfying points can be attributed to dominant co-annihilation channels. As the dark matter mass increases, the thermally averaged cross section decreases; consequently, a smaller mass splitting is required to enhance the co-annihilation rate and achieve the correct relic abundance. The blank region below the colored band for $M_{\rm DM}<700$ GeV corresponds to an under-abundant relic density, primarily due to enhanced co-annihilation cross sections at lower masses. Around $\Delta M\sim$ 50 GeV, the contributions from co-annihilation and annihilation processes become comparable. Above this threshold, annihilation channels dominantly determine the relic density. In this regime, the cross section of Higgs-mediated annihilation processes grows with the Yukawa coupling, $y$, where $y\propto \Delta{M}\sin\theta$. Therefore, with increasing dark matter mass for a fixed $\sin\theta$, a corresponding increase in $\Delta{M}$ is necessary to provide the observed relic abundance. Consequently, the blank region on the right side of the plot represents parameter space yielding an over-abundant relic density while the region to the left is in the correct ball park.
    
Thus we tested the validity of Eq. (\ref{eq:BE1}) for $\sin\theta\in [10^{-4},0.5]$. In other words, solving Eq. (\ref{eq:BE1}) is equivalent to solving Eqs. (\ref{eq:Y1}) and (\ref{eq:Y2}) keeping $\langle\sigma v\rangle_{2010}$ switched off. This confirms the earlier assumption \cite{Bhattacharya:2018fus} that singlet and doublet decouple chemically at the same epoch in the large $\sin\theta$ limit. In this region of parameter space even if we do not switch off the co-scattering term in Eqs. (\ref{eq:Y1}) and (\ref{eq:Y2}), the results remain unchanged. See more discussion below.
    
	\subsubsection{Thermal dark matter relic via the conversion-driven processes in the small $\sin\theta$ regime}\label{sec:co-sactter}
	
	We have already demonstrated in Fig. \ref{fig:case1_anni} ({\it left}) the parameter space for the correct relic of DM in $\sin\theta$-$M_{\rm DM}$ plane, where the singlet and the doublet freeze out at the same epoch through annihilation and co-annihilation processes. Beyond this regime, the singlet may reach equilibrium. However, in this case, the singlet and doublet will decouple at different epochs. Due to small mixing, the singlet will decouple early, while the doublet components will remain in equilibrium for a longer period due to their gauge interactions. Although the singlet decouples early with a larger abundance, the final relic of the DM can be brought to the correct ballpark in the presence of additional conversion-driven processes \cite{DAgnolo:2017dbv, Garny:2017rxs}. 
    
	As discussed in section \ref{sec:annicoanni}, in the large SD mixing angle limit ($\mathcal{O}(10^{-4})<\sin\theta<0.5$) only annihilation and co-annihilation are the dominating processes that decide the DM relic. Had we been added the co-scattering processes to Eqs. (\ref{eq:Y1}) and (\ref{eq:Y2}), the result would not have changed. This can be understood by analyzing Eq. (\ref{eq:gamma21}), where the second term contributes to the effective co-scattering processes. At this point we note that in the large $\sin\theta$ limit, the impact of the co-scattering processes are similar in magnitude to that of the decay and inverse decay processes given in Eq. (\ref{eq:gamma21}). The large co-scattering processes can interchange the sector 1 particles to sector 2 particles efficiently. If the sector 2 particles are already in equilibrium, then they can annihilate to SM particles and deplete the dark matter number density through annihilation and co-annihilation processes $\psi^+\psi^-\rightarrow\rm \rm SM ~SM $ which is independent of the SD mixing angle. On the other hand, if the singlet and doublet decoupled already (the processes 1100,1200 and 2200 decoupled) then the large conversion-driven processes can not alter the total relic, $Y_1+Y_2$, even if these processes remain in equilibrium for a longer period. As a result, the impact of the co-scattering processes is not felt in the large mixing angle limit. In this case, the decoupling of particles in both sectors happens in the same epoch and the Eqs. (\ref{eq:Y1}) and (\ref{eq:Y2}) simplify to the one-sector BE as given in Eq. (\ref{eq:BE1}).

  On the other hand, in the small SD mixing regime, annihilation and co-annihilation processes are sub-dominant. In this limit, Eq. (\ref{eq:Y1}) and (\ref{eq:Y2}) are simplified to
	
	\begin{equation}
		\frac{dY_{1}}{dT}=-\frac{\Gamma_{2\rightarrow 1}}{\mathcal{H}T}\left[Y_{2}-Y_{1}\frac{Y_{2}^{\rm eq}}{Y_{1}^{\rm eq}}\right],
	\end{equation}
	\begin{equation}
		\frac{dY_{2}}{dT}=\frac{s}{\mathcal{H}T}\left[\left<\sigma_{2200}v\right>\left(Y^2_2-{Y^{\rm eq}_2}^2\right)+\frac{\Gamma_{2\rightarrow 1}}{s}\left(Y_{2}-Y_{1}\frac{Y_{2}^{\rm eq}}{Y_{1}^{\rm eq}}\right)\right].
	\end{equation}

\begin{figure}[h]
		\centering
		\includegraphics[scale=0.38]{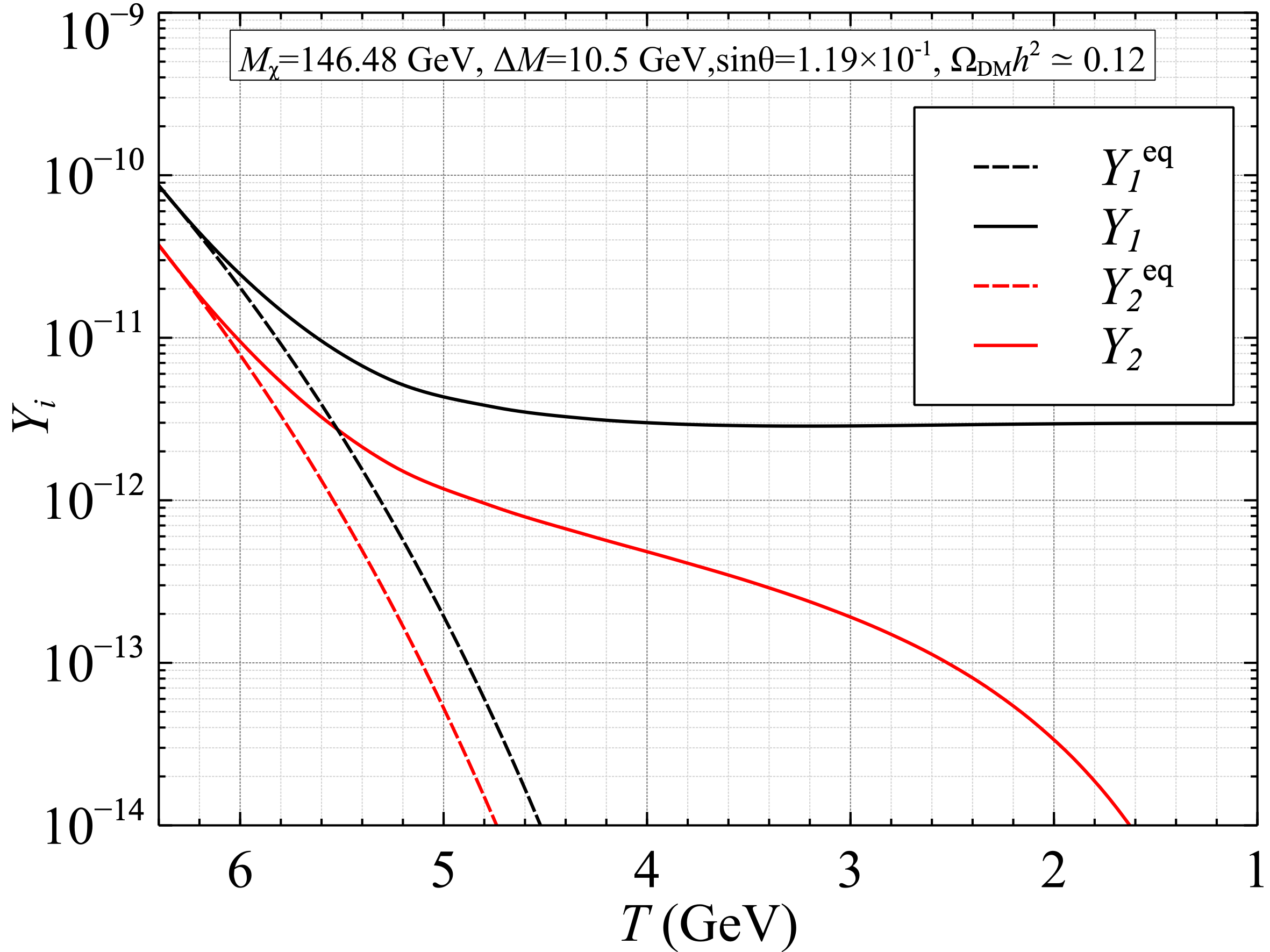}
		\includegraphics[scale=0.38]{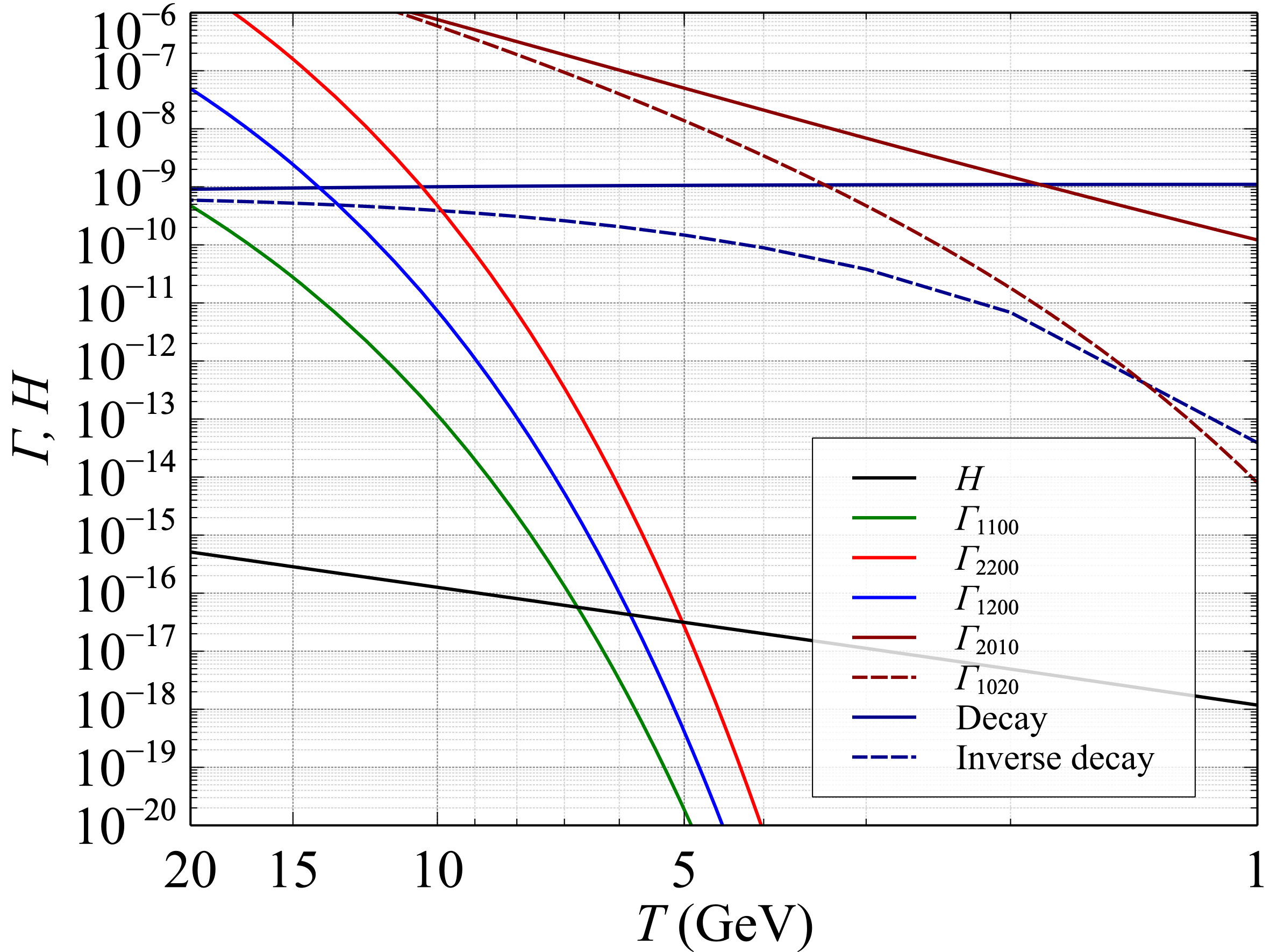}
		\caption{\textit{Left:} Evolutions of sector 1 and sector 2 particle abundances for $M_{\rm DM}$=146.48 GeV, $\Delta M=10.5$ GeV, and $\sin\theta=1.19\times10^{-1}$ are shown for correct DM relic density $\Omega_{\rm DM}h^2\simeq0.12$. \textit{Right:} The comparison of the  interaction rates w.r.t Hubble is shown  as a function of temperature for the parameters given in the \textit{left} panel.}\label{fig:ev1}
\end{figure}
\begin{figure}[h]
		\centering
		\includegraphics[scale=0.38]{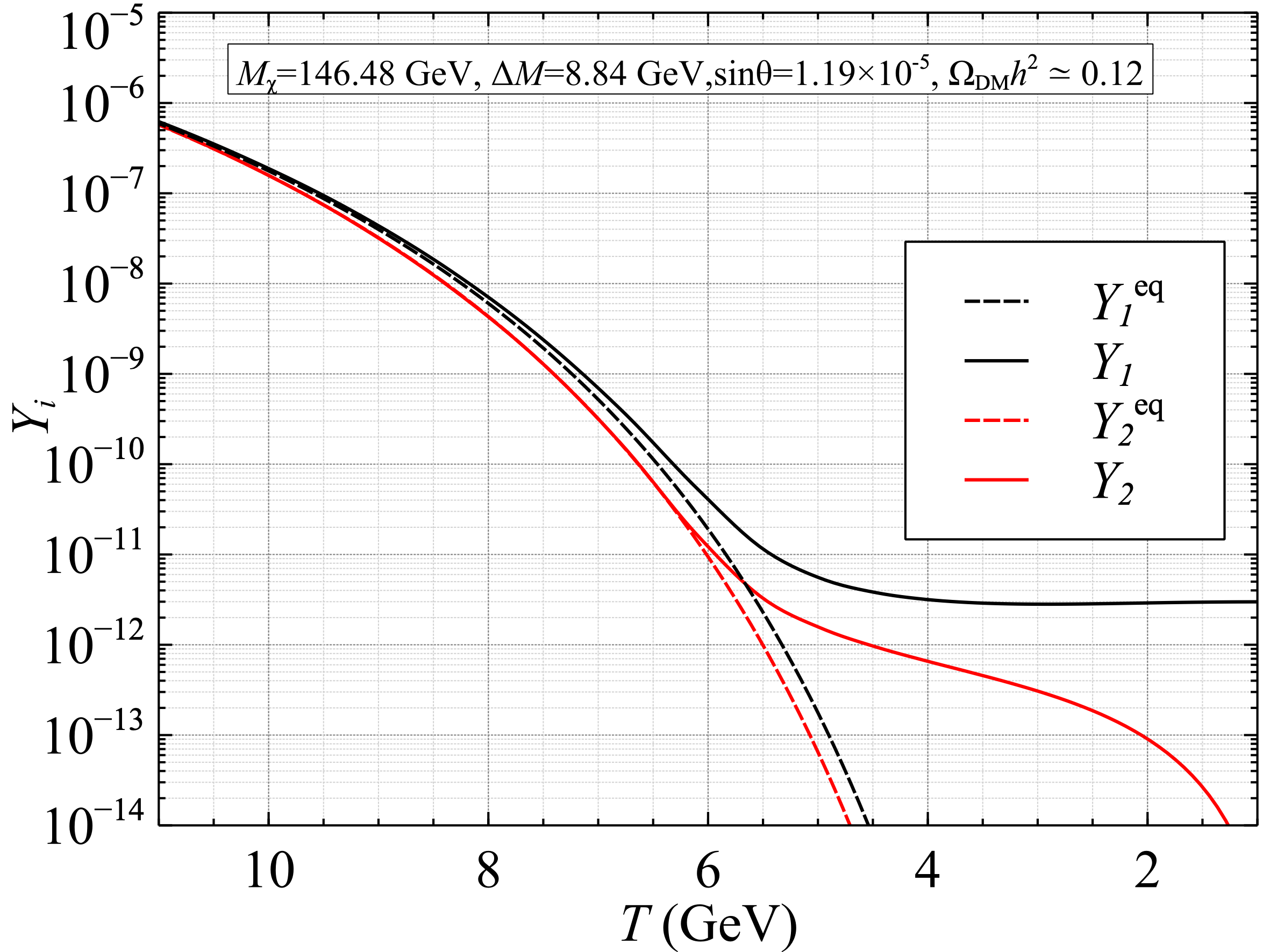}
		\includegraphics[scale=0.38]{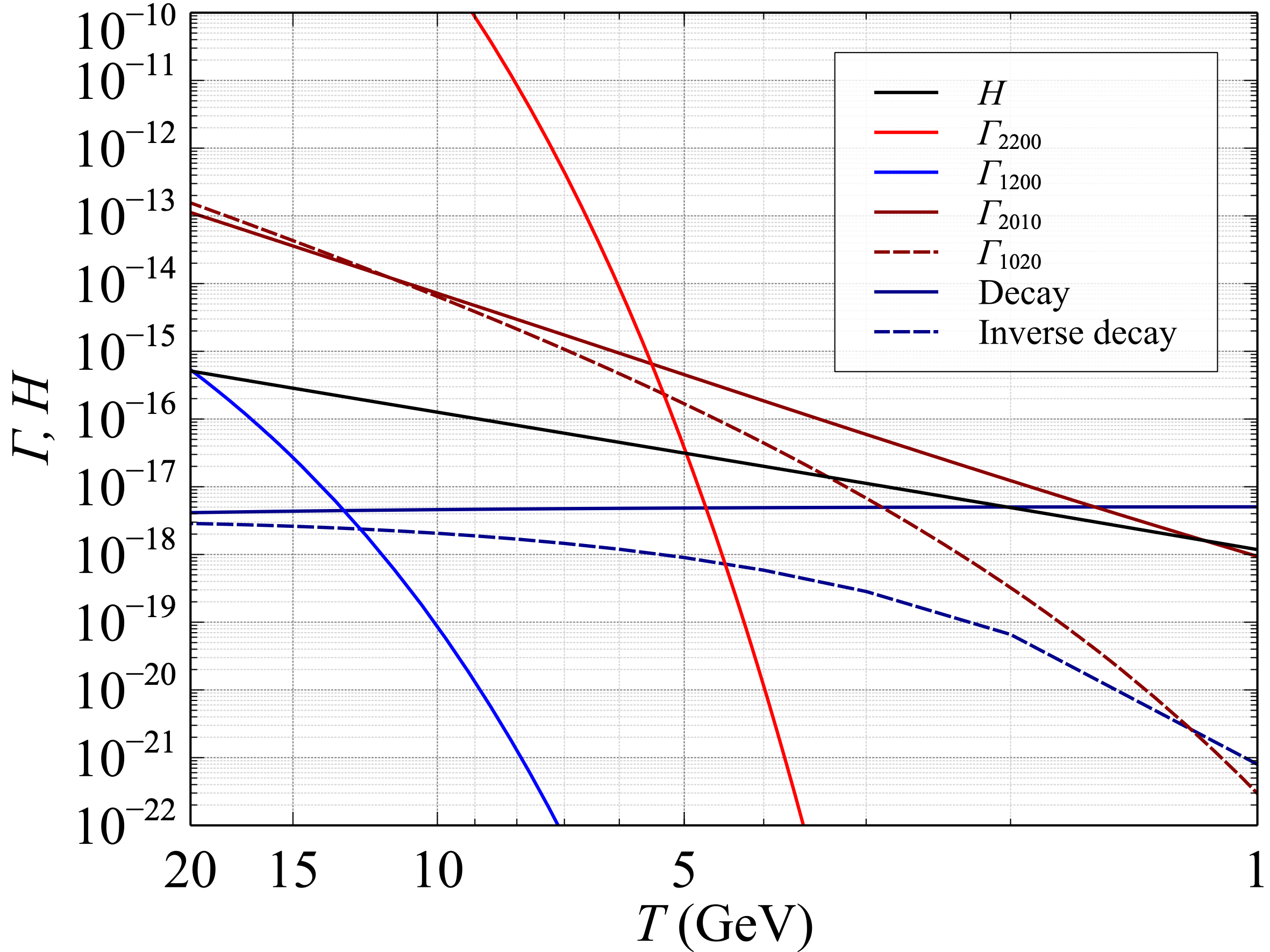}
		\caption{\textit{Left:} Evolutions of sector 1 and sector 2 particle abundances for $M_{\rm DM}$=146.48 GeV, $\Delta{M}=8.4$ GeV, and $\sin\theta=1.19\times10^{-5}$  are shown for correct DM relic density $\Omega_{\rm DM}h^2\simeq0.12$. \textit{Right:}  The comparison of the  interaction rates w.r.t Hubble is shown  as a function of temperature for the parameters given in the \textit{left} panel.}\label{fig:ev2}
\end{figure}

We now explore the regime of the conversion-driven processes,  including co-scatterings, decays and inverse decays, whose impact on the SDDM relic was not studied in earlier works \cite{Bhattacharya:2015qpa,Bhattacharya:2018fus,Bhattacharya:2017sml,Bhattacharya:2018cgx,Bhattacharya:2016rqj,Dutta:2020xwn,Borah:2021khc,Borah:2021rbx,Borah:2022zim,Borah:2023dhk,Paul:2024iie}. We consider two benchmark points BP1, BP2: one in the large mixing limit (BP1 in Fig.~\ref{fig:ev1}) and another in small mixing limit (BP2 in Fig.~\ref{fig:ev2}).\\
\textbf{BP1 ($\sin\theta=0.119$):} In the \textit{left} panel of Fig. \ref{fig:ev1}, we depict the evolution of the abundances of both dark sector species as a function of temperature for mixing angles: $\sin\theta=1.19\times10^{-1}$. The dark matter mass is fixed at $146.48~\mathrm{GeV}$, and $\Delta{M}=10.5~\mathrm{GeV}$, selected to obtain the correct relic abundance. The comparison among relevant interaction rates is displayed in the \textit{right} panel of Fig.~\ref{fig:ev1}. Here, we observe that the annihilation rate of the singlet component (1100 processes) falls below the Hubble rate first, followed by the decoupling of co-annihilation processes (1200 processes). Both sectors remain in chemical equilibrium down to $T\sim6~\mathrm{GeV}$, as seen in the \textit{left} panel of Fig.~\ref{fig:ev1}. On the other hand, the rate of conversion-driven processes (decay, inverse decay and co-scattering) remain significantly larger than the Hubble rate during this period. These processes can efficiently deplete the number density of dark matter by efficiently converting the sector 1 particles to sector 2 particles followed by annihilation to SM particles (2200 processes). Once the 2200 processes decouple (shown by the red line in \textit{right} panel of Fig.~\ref{fig:ev1}) around $T\sim5$ GeV, the number density is settled to its final value as seen in the \textit{left} panel of Fig.~\ref{fig:ev1}.\\
\textbf{BP2 ($\sin\theta=1.19\times10^{-5}$):} In the \textit{left} panel of Fig.~\ref{fig:ev2}, we depict the evolution of the abundances of both dark sector species as a function of temperature for mixing angles: $\sin\theta=1.19\times10^{-5}$. The DM mass is taken to be 146.48 GeV, and $\Delta{M}=8.84~\mathrm{GeV}$ to get the correct relic density. Since the rate of sector 1 annihilation processes (1100) is proportional to $\sin^4\theta$, these processes are negligible in comparison to co-annihilation processes (1200) for this BP as the rate of co-annihilation processes are proportional to $\sin^2\theta$. As shown in the \textit{right} panel of Fig.~\ref{fig:ev2}, the co-annihilation processes decouple at approximately $T\sim20~\mathrm{GeV}$. As a result the chemical decoupling of sector 1 particles happens with a relatively larger abundance, while sector 2 particles remain in equilibrium due to their gauge interactions, as shown in the \textit{left} panel of Fig.~\ref{fig:ev2}. Note that in the small mixing angle limit, the decay and inverse decay processes are subdominant in comparison to co-scattering processes. This is evident from the right panel of Fig.~\ref{fig:ev2}. It is worth mentioning that, in this case, processes involving light SM fermions contribute to the co-scattering processes significantly, \textit{e.g.} $\chi_1e^-\rightarrow\psi^-\nu_e,\chi_1\nu_e\rightarrow\psi^+e^-$ etc. In the presence of efficient co-scattering, sector 1 particle can upscatter with SM particles into sector 2, followed by annihilation to SM via 2200 processes. Once the 2200 processes decouple, the relic settles to its final value even if the rate of co-scattering processes continue to be in equilibrium for a longer period. For this BP the 2200 processes decouple at $T\sim5$ GeV (see red line in the \textit{right} panel of Fig.~\ref{fig:ev2}.). As a result the relic settles around $T\sim5$ GeV as seen from the \textit{left} panel of Fig.~\ref{fig:ev2}.

\begin{figure}[h]
		\centering    
		\includegraphics[scale=0.38]{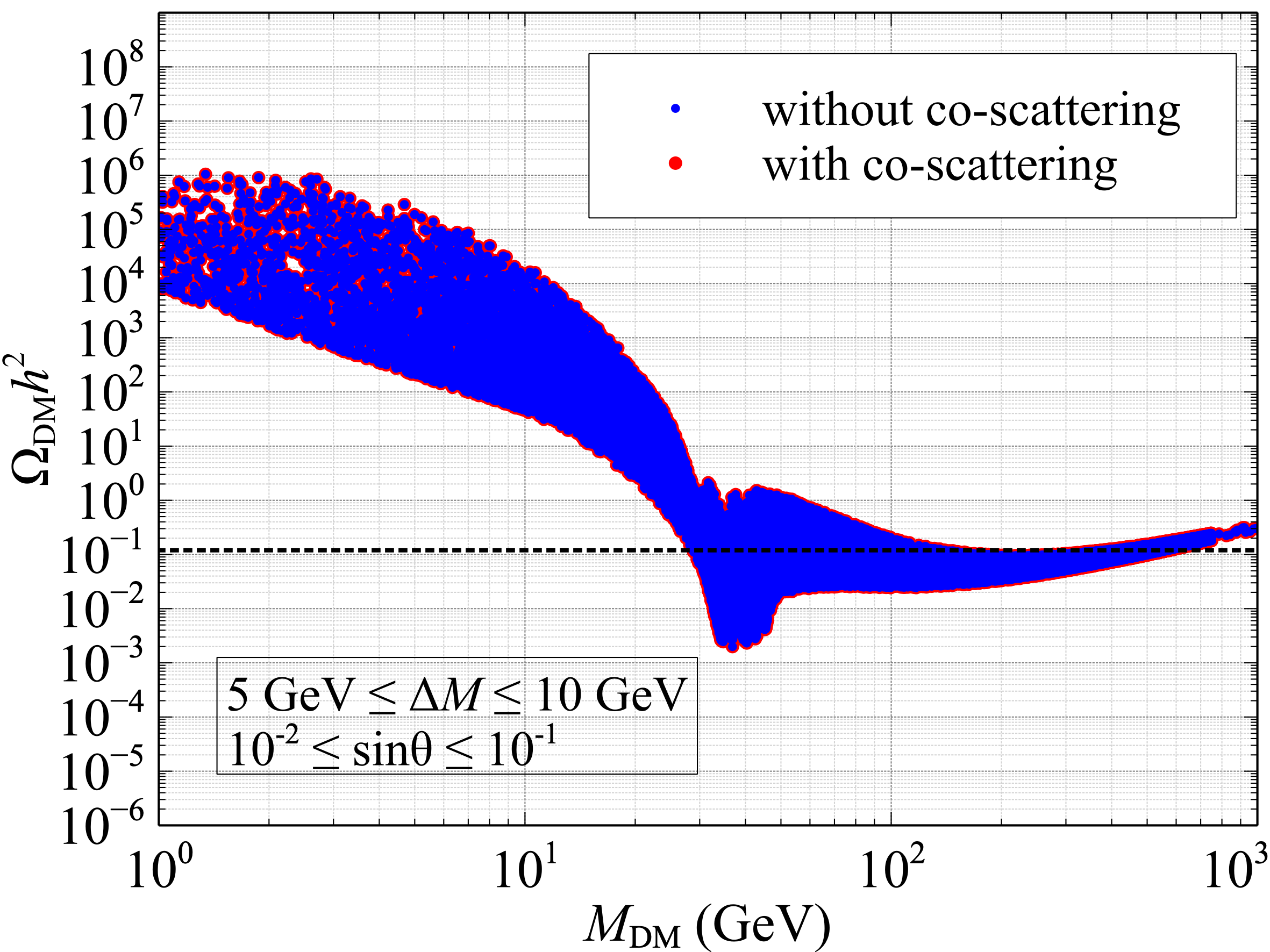}
		\includegraphics[scale=0.38]{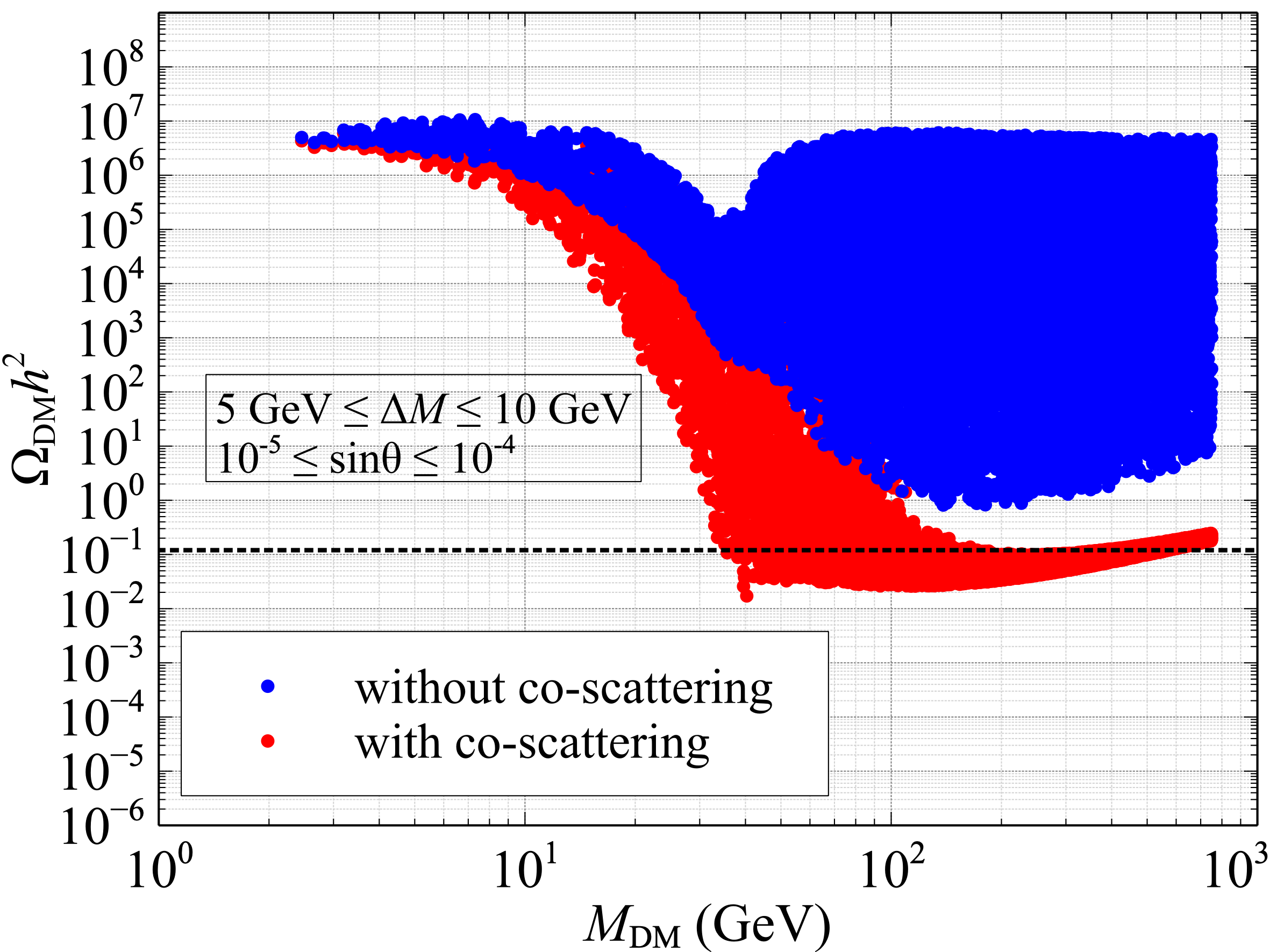}
		\caption{Variation of DM relic density w.r.t DM mass with the fixed mass splitting of $5~{\rm GeV}\le\Delta{M}\le10$ GeV for $10^{-2}\le\sin\theta\le10^{-1}$ [{\it left}] and $10^{-5}\le\sin\theta\le10^{-4}$ [{\it right}]. For the blue colored points, we have solved Eqs. (\ref{eq:Y1}) and (\ref{eq:Y2}), by switching off  the co-scattering processes, while for the red points we have considered all the processes given in Eqs. (\ref{eq:Y1}) and (\ref{eq:Y2}).}
		\label{fig:parameterspace2}
	\end{figure}

To explore in details   the impact of co-scattering on the relic density of DM, we performed several analyses and examined the allowed parameter space by solving Eq. (\ref{eq:Y1}) and (\ref{eq:Y2}).
\begin{figure}[h]
		\centering        \includegraphics[scale=0.38]{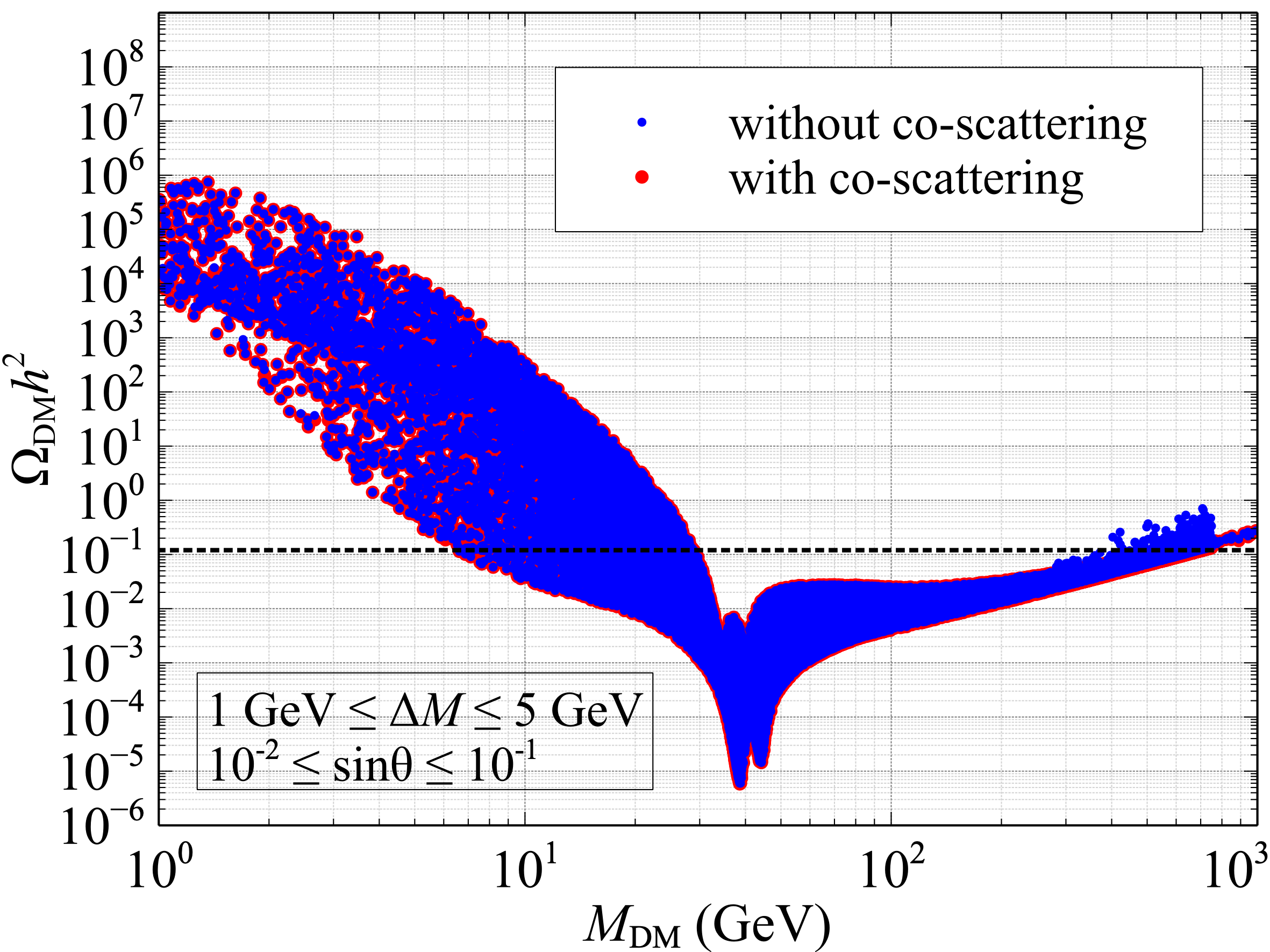}
		\includegraphics[scale=0.38]{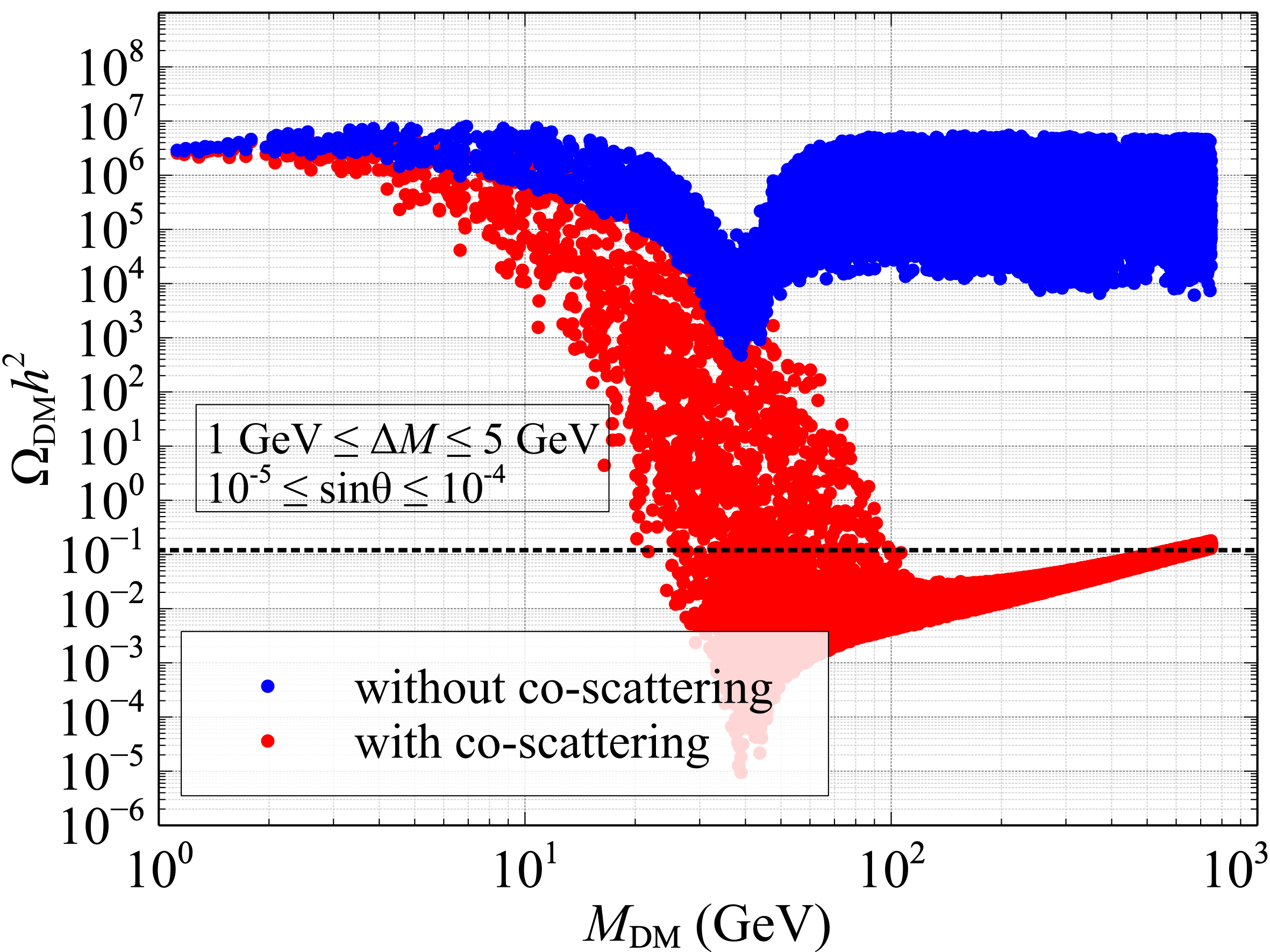}
		\caption{Variation of DM relic density w.r.t. DM mass with the fixed mass splitting of $1~\rm GeV\le\Delta{M}\le5$ GeV for $10^{-2}\le\sin\theta\le10^{-1}$ [{\it left}] and $10^{-5}\le\sin\theta\le10^{-4}$ [{\it right}]. For the blue colored points, we have solved Eqs. (\ref{eq:Y1}) and (\ref{eq:Y2}), by switching off  the co-scattering processes, while for the red points we have considered all the processes given in Eqs. (\ref{eq:Y1}) and (\ref{eq:Y2}).}
		\label{fig:parameterspace1}
	\end{figure}
In Fig. \ref{fig:parameterspace1}, we have kept the SD mass splitting in the range of 1 GeV to 5 GeV and then varied  $\sin\theta$ in two sets. In the {\it left} panel of Fig. \ref{fig:parameterspace1}, $\sin\theta$ is varied in the range $10^{-2}-10^{-1}$, while in the \textit{right} panel it is varied in the range $10^{-5}-10^{-4}$ . In the \textit{left} panel, due to smaller mass splitting and the large mixing angle, the co-annihilations are the most dominant processes that determine the relic of the DM. We see that the red points (obtained by solving Eqs. (\ref{eq:Y1}) and (\ref{eq:Y2})) and blue points (obtained by solving Eq (\ref{eq:Y1}) and (\ref{eq:Y2}), but keeping the co-scattering processes off) coincide with each other\footnote{In Fig. \ref{fig:parameterspace1}, and Fig. \ref{fig:parameterspace2} \textit{without co-scattering} points are obtained using the function {\tt darkOmegaN} with the flag {\tt ExcludedForNDM="2010"}, whereas \textit{with co-scattering} points are obtained using {\tt darkOmegaN} function without any flag.}. This implies that in the large $\sin\theta$ limit, the impact of co-scattering is not felt. This is because, for large $\sin\theta$, the sector 1 (singlet) and sector 2 (doublet) decouple at the same epoch. In other words, the processes 1100, 1200 and 2200 decouple close to each other. In this case, even if the rate of co-scattering processes are large, they are not affecting the overall relic abundance ($Y_1+Y_2$) as the sector 1 and sector 2 particles are already decoupled from the thermal bath. In this case, the conversion-driven processes can at most interchange the number density of the sector 1 and sector 2 particles. However, the total dark sector relic, $Y_1+Y_2$ will be remain unchanged.
    
In the \textit{right} panel of Fig. \ref{fig:parameterspace1}, because of the small mixing angle ($10^{-5}\le\sin\theta\le10^{-4}$), the annihilation/co-annihilation processes become subdominant. In the absence of co-scattering processes, the DM gets chemically decoupled with a larger DM abundance as shown by the blue points. When we include the co-scattering effects, the dynamics of the freeze-out of DM change drastically. In particular these processes further deplete the DM relic density by efficiently converting the sector 1 particles to sector 2 particles followed by annihilation to SM bath until the 2200 processes are decoupled. This is shown by the red color points.
We see that the co-scattering effects are significant in the small mixing angle regime while their effects are not prominent for the large mixing angles even if the rate of these processes are large.
	
We now move to Fig. \ref{fig:parameterspace2}, where we have varied the mass splitting in the range from 5 GeV to 10 GeV, while the SD mixing is kept in the same range as in Fig. \ref{fig:parameterspace1}. We observe that the dependency of the DM relic on the SD mixing angles remains similar as in Fig. \ref{fig:parameterspace1}. When we increase the mass splitting, the co-annihilation and co-scattering crosssection becomes smaller, which leads to larger DM abundance in comparison to Fig. \ref{fig:parameterspace1}. This is clearly visible from the {\it left} panel of Fig. \ref{fig:parameterspace2}. Now we compare the red colored points in the right panels of Fig. \ref{fig:parameterspace1} and \ref{fig:parameterspace2} where the $\Delta{M}$ increases in Fig. \ref{fig:parameterspace2} compared to Fig. \ref{fig:parameterspace1} while all other parameters remain same. A larger $\Delta M$ reduces the co-annihilation and co-scattering effects. Moreover, in the small $\sin\theta$ limit co-scattering effects always dominate over decay inverse decay. As a result we get an overall increment in the dark matter relic abundance. This can be easily seen by looking at the red colored points in the right panel of Fig. \ref{fig:parameterspace1} and \ref{fig:parameterspace2}. We then compare the blue colored points (co-scattering effects are not included while decay and inverse decay terms are present, see Eq. (\ref{eq:gamma21})) in the right panel of Fig. \ref{fig:parameterspace1} and \ref{fig:parameterspace2}. As the $\Delta{M}$ increases in Fig. \ref{fig:parameterspace2} w.r.t Fig. \ref{fig:parameterspace1}, the co-annihilation effects decrease. On the other hand, the decay and inverse decay effects increase as they are proportional to $(\Delta{M})^5$ (see Eq. (\ref{eq:3bodydecay})). The latter processes help in depleting the relic for a longer period. As a result we get an overall reduced relic even though there is a small  relic enhancement due to the reduced co-annihilation effect.
	\begin{figure}[h]
		\centering  
		\includegraphics[width=7.5cm,height=7.5cm]{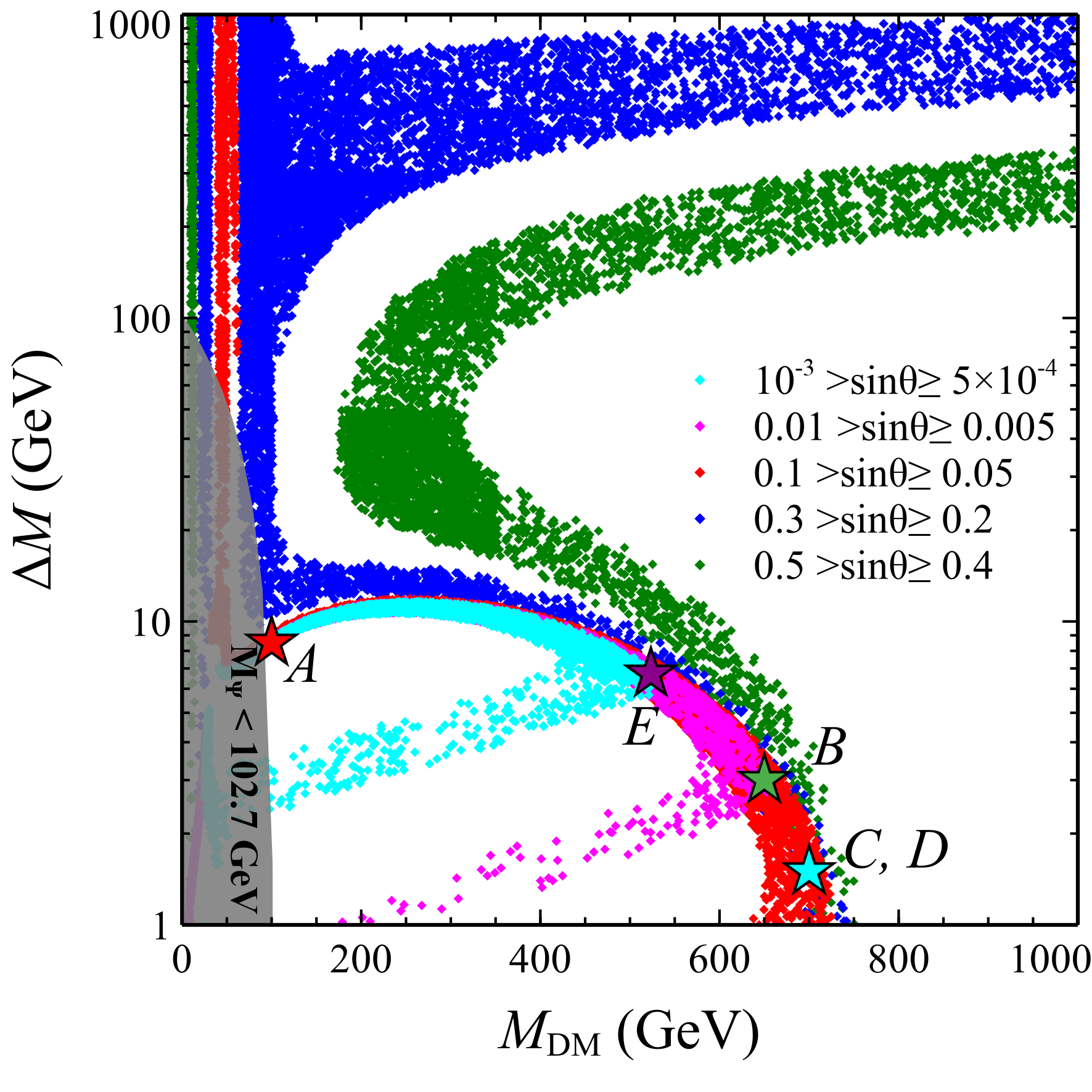}  
		\includegraphics[width=7.5cm,height=7.5cm]{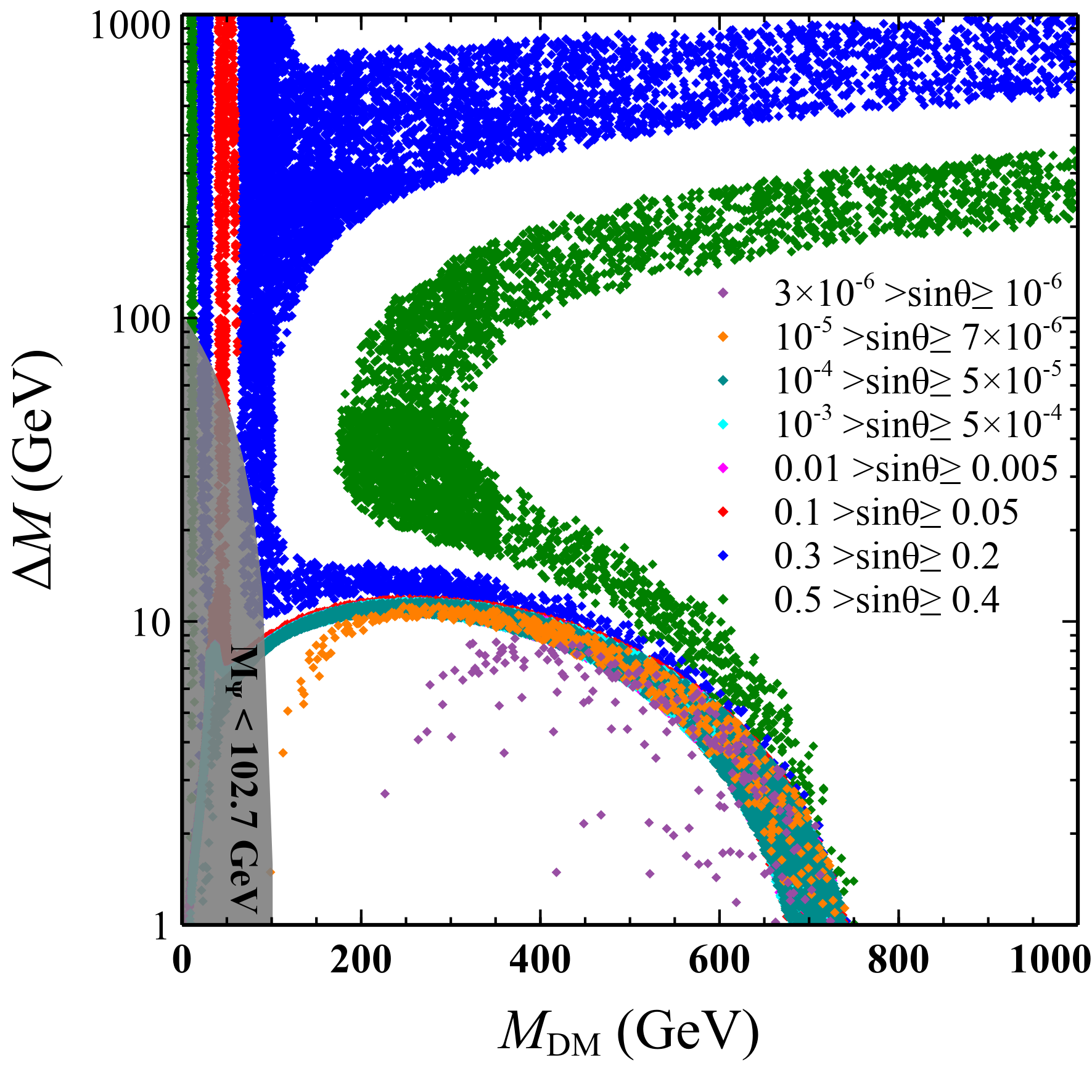}
		\caption{
    \textit{Right:} The correct relic density is shown in the plane $\Delta{M}-M_{\rm DM}$ with different ranges of mixing angles as provided in the figure inset by solving Eqs. (\ref{eq:Y1}) and (\ref{eq:Y2}). \textit{Left:} Same as in the \textit{right} panel, except that the co-scattering effects are switched off.}
		\label{fig:cosscattering04}
	\end{figure}
	
In Fig. \ref{fig:cosscattering04}, we present the points that yield the correct DM relic density in the $\Delta M$-$M_{\rm DM}$ plane by solving Eqs. (\ref{eq:Y1}) and (\ref{eq:Y2}). In the {\it left} plot, we show the points that give the correct relic density when co-scattering processes are switched off. In this case, the relic density calculation considers only annihilation, co-annihilation, decay and inverse decay processes. The color coding represents different values of $\sin\theta$, ranging from 0.5 to $5\times10^{-4}$, as indicated in the plot. For $\sin\theta \ge 0.05$, the singlet and doublet components decouple almost at the same epoch. Therefore, our results, obtained by solving Eqs. (\ref{eq:Y1}) and (\ref{eq:Y2}), are consistent with previous findings \cite{Bhattacharya:2018fus}, obtained by solving Eq (\ref{eq:BE1}) and depicted in the right panel of Fig. \ref{fig:case1_anni}. For $\sin\theta < 0.05$, singlet and doublet components decouple at different epochs. As a result we find new parameter space shown in magenta and cyan colored points in comparison to the right panel of Fig. \ref{fig:case1_anni}.

To explain our findings, we identify five sample points: $A, B, C,D$ and $E$ as shown in the $left$ panel of Fig. \ref{fig:cosscattering04}. For a typical SD mixing, represented by magenta-colored points, the cross-section decreases as we move from left to right due to the increasing DM mass. This decrement can be compensated by reducing $\Delta M$. In absence of decay and inverse decay terms in Eqs. (\ref{eq:Y1}) and (\ref{eq:Y2}), for larger $\Delta M$ [e.g., point $A$], the co-annihilation cross-section experiences greater suppression, causing DM to decouple earlier and resulting in a larger abundance. However, with decay terms ($\Gamma_{\Psi\rightarrow\chi,\rm SM}$) included, the DM relic density can be brought to the correct range. Table \ref{tab:tab1} shows that for point $A$, the DM is overabundant in absence of decay term\footnote{In Table \ref{tab:tab1}, last column corresponds to the relic without considering co-scattering and decay, which we denote as  \textit{$\Omega_{2s}h^2$(no co-scattering) without decay}. This can be achieved using the function {\tt darkOmegaN} with the flag {\tt ExcludedForNDM="2010 DMdecay"}. On the other hand \textit{$\Omega_{2s}h^2$(no co-scattering)} is obtained using the function {\tt darkOmegaN} with the flag {\tt ExcludedForNDM="2010"}.}. In presence of the decay and inverse-decay processes in  the Boltzmann equations (\ref{eq:Y1}) and (\ref{eq:Y2}), sector 1 to sector 2 conversion can happen efficiently for a longer duration, thereby reducing the relic abundance to the desired value. The evolution of yields ($Y_1$ and $Y_2$) both considering decay and without decay has been shown in Fig. \ref{fig:ev_point_A} in Appendix \ref{app:evo_plot}. In absence of the decay terms, a relatively small $\Delta M$ [e.g., point $B$] delays the DM decoupling epoch, resulting in a smaller yield in comparison to point $A$ ($Y_{1}(B) < Y_{1}(A)$). In this scenario, a smaller contribution from decay and inverse-decay processes is sufficient to achieve the correct relic density. This is evident from Table \ref{tab:tab1}, where turning off the decay term results in a smaller increase in relic density compared to point $A$. For further smaller $\Delta M$ [e.g., point $C$ where $\sin\theta=9.5\times10^{-3}$], co-annihilation combined with decay-inverse decay fail to maintain equilibrium, leading to DM freezing out with a larger abundance in comparison to point $B$. This behavior can be understood from the decay term: $\Gamma_{\Psi\rightarrow\chi,\rm SM} \propto \sin^2\theta \Delta M^5$. Since $\sin\theta$ is same for point $B$ and $C$, the reduced $\Delta M$ at the point $C$ gives a smaller $\Gamma$, thereby resulting a larger relic. By increasing $\sin\theta$ from $9.5\times10^{-3}$ to $9.9\times10^{-2}$ we recover the correct relic density as shown in Table \ref{tab:tab1}.

For smaller mixing angles, typically $\sin\theta<0.05$, we have new sets of point that satisfy correct relic density. In particular, the set of points represented by magenta and cyan color curve back leftward from point $B$ and $E$, respectively. In this regime, decreasing DM mass and $\Delta M$, increase the cross-section, leading to a late decoupling of DM with a comparatively smaller abundance yet larger than the required relic. In this case, the small contribution from decay terms are required to give rise the correct relic density. The position of the turning points $B$ and $E
$ varies with $\sin\theta$. For smaller SD mixing, it occurs at smaller DM masses, while for larger SD mixing, it occurs at larger DM masses. This can be understood as follows. The cross section usually decreases with increase in mass, leading to a larger chemically decoupled relic. Therefore, for a larger DM mass, we need a larger $\sin\theta$ to bring down the relic to the correct ballpark and vice-versa.
	
	\begin{table}[h]
		\centering
		\resizebox{15cm}{!}{
			\begin{tabular}{|c|c|c|c|c|c|}
				\hline\hline
				Point & $M_{\rm DM}$ (GeV) &$\Delta{M}$ (GeV) & $\sin\theta$ & $\Omega_{2s}h^2$(no co-scattering)& $\Omega_{2s}h^2$(no co-scattering) without decay\\
				\hline
				$A$&100&8.5 & $6\times10^{-3}$&0.1128 & $4.795\times10^{2}$\\
				\hline
				$B$ & 650 & 3& $9.5\times10^{-3}$&0.1269 &$2.503\times10^1$ \\
				\hline
				$C$& 700 &1.5 &$9.5\times10^{-3}$ &0.3445 & $2.114\times10^1$ \\
				\hline
				$D$& 700 &1.5 &$9.9\times10^{-2}$ &0.1234 & 0.1434\\
                \hline
				$E$& 520 &6.2 &$10^{-3}$ &0.1247 &$ 2.23\times 10^{3}$\\
				\hline\hline
			\end{tabular}
		}
		\caption{Relic density for benchmark points: (a) without considering co-scattering, and (b) without considering co-scattering and decay.}
		\label{tab:tab1}
	\end{table}
	
As discussed earlier, for larger SD mixing ($\sin\theta\gtrsim0.05$) the sector 1 (singlet) and sector 2 (doublet) decouple almost at the same epoch. In this case, the co-scattering effects does not play vital role in deciding the relic density of DM, even though the rate of co-scattering processes are large. This implies our results are consistent with previous findings \cite{Bhattacharya:2018fus}. However, for smaller SD mixing ($\sin\theta<0.05$), the co-scattering effects are important.
In the {\it right} panel of Fig. \ref{fig:cosscattering04}, we present the correct relic density points by solving the Boltzmann equations (\ref{eq:Y1}) and (\ref{eq:Y2}), without excluding the contribution from co-scattering processes. The parameter space that previously yielded the correct relic density when co-scattering was excluded get under abundant in presence of the co-scattering processes. Conversely, the over-abundant points with $\sin\theta < 10^{-4}$, which did not provide the correct relic density in the absence of co-scattering processes, now yield the correct relic density once these processes are included. This can be understood by  examining Eq. (\ref{eq:gamma21}), which includes both co-scattering as well as decay and inverse decay processes. In the small $\sin\theta$ limit, the co-scattering rates are much larger than the decay and inverse decay rates (see \textit{right} panel of Fig. \ref{fig:ev2}), even though they have same $\sin\theta$ dependency. Therefore, for a given $\sin\theta$, which yields the correct relic density in the absence of co-scattering, become under abundant in presence of co-scattering processes as the latter processes will convert the sector 1 to sector 2 particles followed by annihilation to SM bath in a faster rate. We consolidate our claim by defining a new parameter, $\rho_2$, as
	\begin{equation}
		\rho_{2}=\frac{\Omega_{2s}h^2}{\Omega_{2s}h^2({\rm no~coscattering})},\label{eq:rho2}
	\end{equation}
where $\Omega_{2s}h^2$ represents relic density obtained by solving Eqs. (\ref{eq:Y1}) and (\ref{eq:Y2}), while $\Omega_{2s}h^2$({\rm no~co-scattering}) denotes the solution of Eqs.(\ref{eq:Y1}) and (\ref{eq:Y2}) in the absence of the co-scattering contribution. Eq. (\ref{eq:rho2}) implies that, $\rho_2=1$ indicates the relic density is independent of co-scattering effects. Consequently, any deviation from 1 will indicate the role of co-scattering in the final relic density.
The larger is the deviation, the more is the effect from co-scattering. In fact, $\Omega_{2s}h^2$ includes all the processes involving annihilation, co-annihilation, co-scattering, and decay, while $\Omega_{2s}h^2$ (no co-scattering) involves only annihilation, co-annihilation, and decay. This implies the effective cross-section, giving rise to $\Omega_{2s}h^2$, is always larger than or equal to $\Omega_{2s}h^2$ (no co-scattering). Therefore, $\Omega_{2s}h^2 \le\Omega_{2s}h^2$ (no co-scattering), resulting in $\rho_2\le1$.	
	\begin{figure}[h]
		\centering        \includegraphics[scale=0.5]{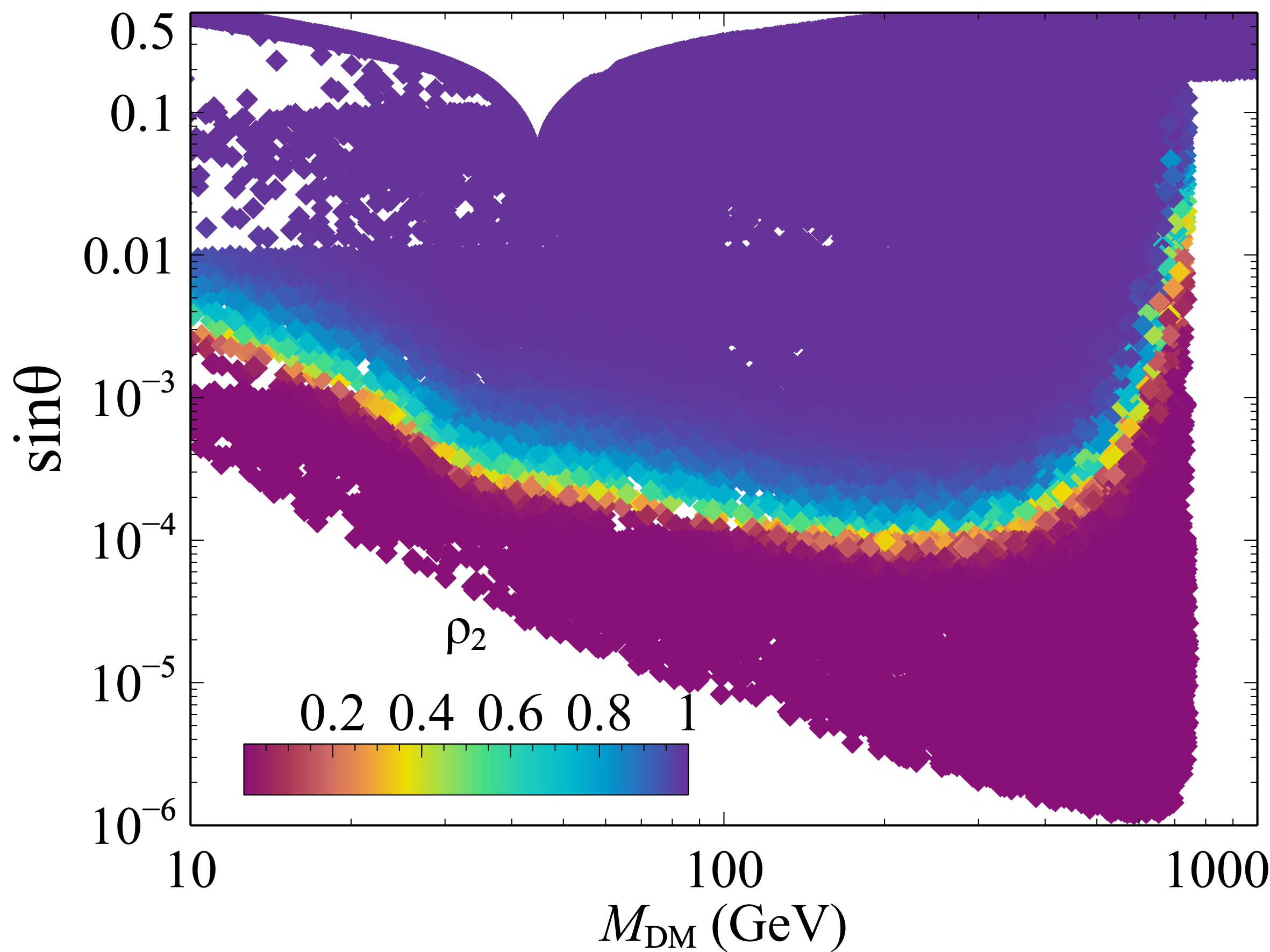}
		\caption{The co-scattering parameter space in the plane of $\sin\theta$ vs $M_{\rm DM}$. The color code represents the parameter $\rho_{2}$.}
		\label{fig:parameterspace3}
	\end{figure}
	
In Fig. \ref{fig:parameterspace3}, we present all the points that satisfy correct relic density in the plane of $\sin\theta$ vs $M_{\rm DM}$. The range of $\sin\theta$ is chosen in such a way that the DM will be in equilibrium. The color code depicts the $\rho_2$ values. If $\rho_2$=1, then the effect of co-scattering is negligible, which is shown by the violet-colored points. As we decrease the value of $\sin\theta$, the $\rho_2$ deviates from 1, which implies the nontrivial role of co-scattering in large parameter space shown by cyan, yellow, and purple color points.
Below this region, the DM is produced thermally. However the conversion-driven processes (both decay, inverse decay and co-scattering) are not efficient enough due to further smaller $\sin\theta$. As a result, in this region, the DM remains over abundant. Beyond certain threshold limit of SD mixing, the DM (dominantly singlet component) never comes to thermal equilibrium. In this region of parameter space, the DM can be produced via freeze-in processes as we discussed below.

\subsection{Dark matter production via freeze-in mechanism}\label{sec:freezein}

In Sections \ref{sec:annicoanni}, and \ref{sec:co-sactter}, we saw that the singlet is brought to equilibrium for the $\sin\theta\gtrsim\mathcal{O}(10^{-7})$ depending on the DM mass. This implies that for $\sin\theta\gtrsim\mathcal{O}(10^{-7})$, the relic density of SDDM will be obtained via the freeze-out mechanism. If the SD mixing is further reduced, the DM relic density can no longer be determined through the conventional freeze-out mechanisms. However, it can be produced through the freeze-in mechanism. In this regime, we take into account all the $2\rightarrow2$ processes along with decay and inverse decay processes and solve Eq. (\ref{eq:Y1}) and (\ref{eq:Y2}) with initial conditions, $Y_2=Y_2^{\rm eq}$ and $Y_1=0$. To estimate the final DM relic density, we used $\Omega_{\rm DM}h^2=0.12\left(\frac{(Y_1+Y_2)_{T\rightarrow 0}}{4.26\times 10^{-10}}\right)\left(\frac{M_{\rm DM}}{{1~\rm GeV}}\right)$.
\begin{figure}[h]
		\centering    
		\includegraphics[scale=0.5]{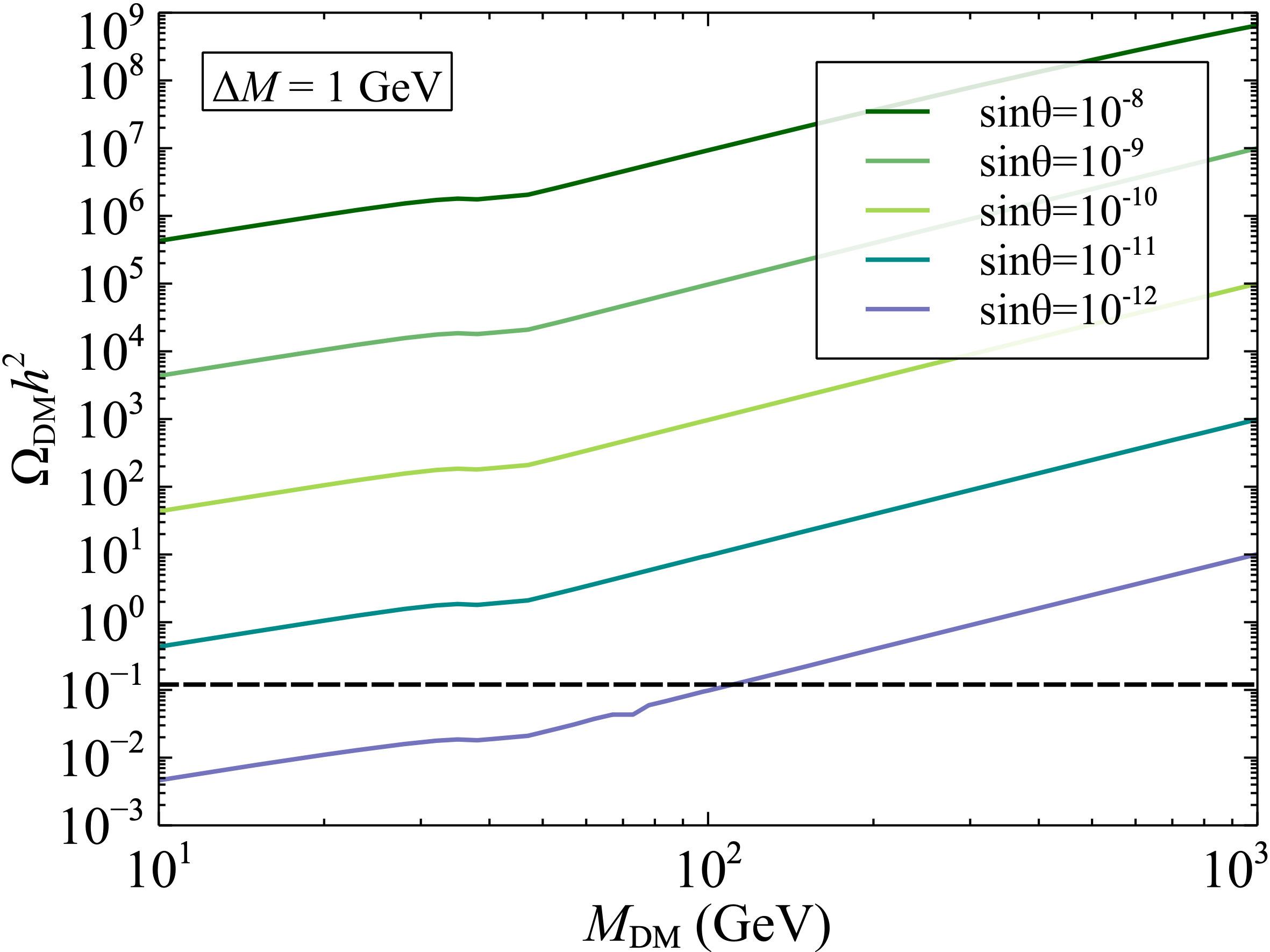}
		\caption{Variation of relic density with DM mass for a fixed $\Delta{M}=1$ GeV. We take 5 values of $\sin\theta$ as shown by different colors. The black dashed line represents the required relic density.}
		\label{fig:freezeinscan}
	\end{figure} 
	\begin{figure}[h]
		\centering    
		\includegraphics[scale=0.5]{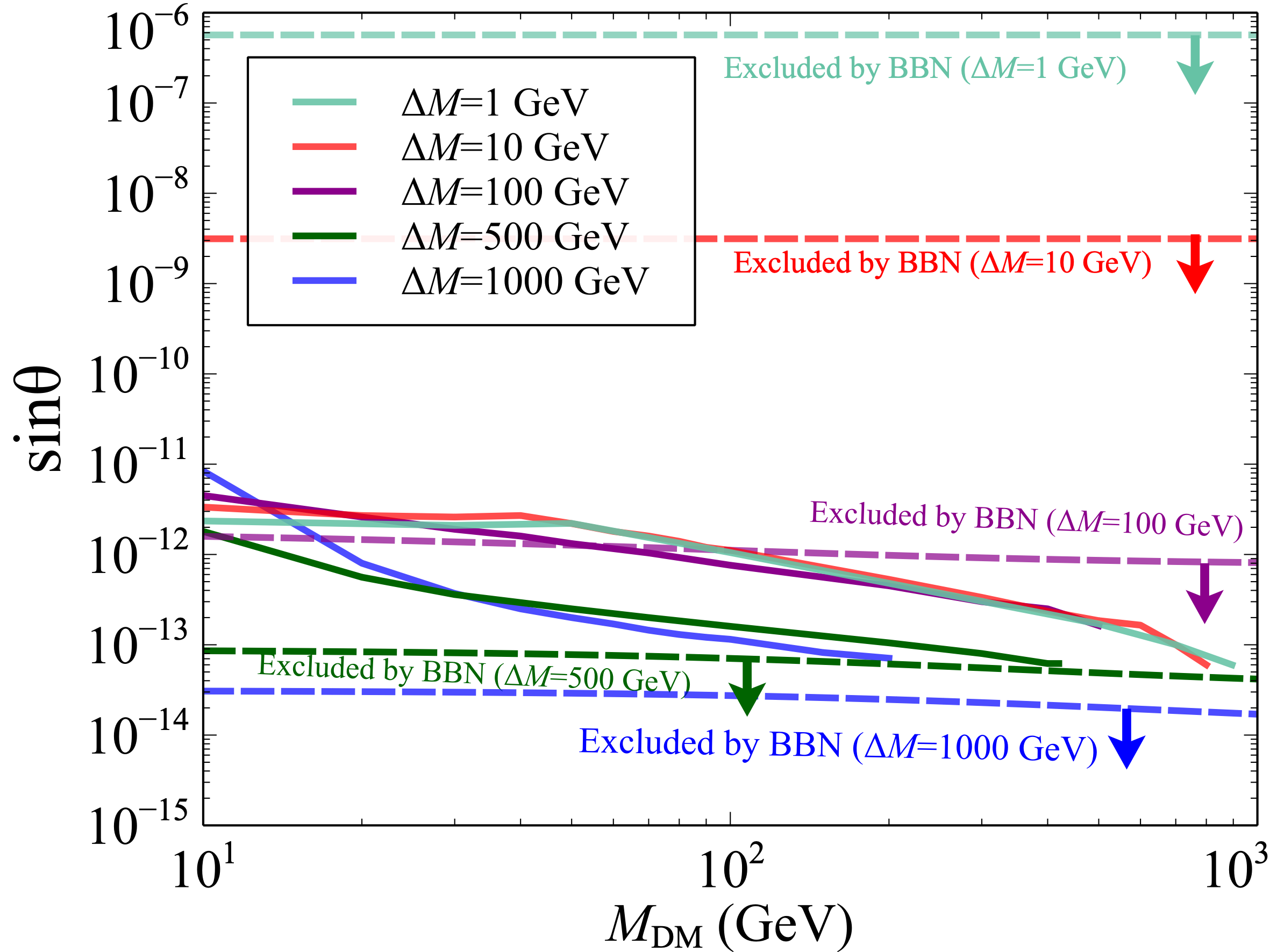}
		\caption{Contours of the correct relic density obtained through freeze-in processes are shown by the colored solid lines in the plane of $\sin\theta$ vs DM mass. The corresponding BBN constraints are shown by colored dashed lines.}
		\label{fig:freezein}
	\end{figure}
In Fig. \ref{fig:freezeinscan}, we present discrete contours of $\sin\theta$ in the $\Omega_{\rm DM}h^2- M_{\rm DM}$ plane, assuming a fixed SD mass splitting $\Delta M=1$ GeV. We observe that increasing $\sin\theta$ enhances the rates of both $2\rightarrow2$ scattering and decay-induced production processes, leading to a higher dark matter relic abundance. Additionally, for each $\sin\theta$ the relic density increases with increasing $M_{\rm DM}$. This behavior arises because, in the freeze-in production scenario, the dark matter yield saturates around $T_{\rm FI}\sim M_{\rm DM}$. A higher freeze-in temperature $(T_{\rm FI})$ corresponds to a larger production rate, thereby resulting in a larger relic abundance for larger $M_{\rm DM}$.

In Fig. \ref{fig:freezein}, we have shown the correct relic density contours with solid lines for five different mass splittings, $\Delta{M}=1,10,100,500,1000$ GeV, in the plane of $\sin\theta$ verses $M_{\rm DM}$. We also show the corresponding BBN constraints\footnote{We take a conservative time scale of $\tau_{BBN}=1$ s. In the region below the dashed lines, $\tau_{BBN}<\tau_{\psi}$, and hence ruled out.} with dashed lines, arising from the doublet decay to the singlet. Since these contours (solid lines) correspond to correct relic density, with increase in mass, the number density is expected to decrease\footnote{As $\Omega_{\rm DM}h^2\propto M_{\rm DM}n_{\rm DM}$, with increase in mass of DM, the number density of DM should decrease to provide relic density in the correct ballpark. It is possible if we decrease the $\sin\theta$ which leads to a reduction in the DM number density.}, which requires a smaller $\sin\theta$. This behavior can be easily read from the solid contours shown in Fig. \ref{fig:freezein}. When $\Delta M\lesssim80$ GeV, the freeze-in relic density is primarily determined by three-body decay modes of the doublet along with other $2\rightarrow2$ processes like $22\rightarrow11,20\rightarrow10$, etc. Conversely, as $\Delta M$ exceeds 80 GeV, 91 GeV, and 125 GeV, the two-body decay channels $\psi^\pm\rightarrow\chi_1 W^\pm$, $\chi_0\rightarrow\chi_1 Z$, and $\chi_0\rightarrow\chi_1 H$ become accessible along with the above $2\rightarrow2$ processes. The red solid line represents $\Delta{M}=10$ GeV. The corresponding BBN constraint excludes this $\Delta{M}$ for all range of DM masses. On the other hand, BBN excludes $M_{DM}>60$ GeV for $\Delta{M}=100$ GeV which is shown with the magenta colored line. The green colored line represents $\Delta{M}=500$ GeV. We see that the correct relic can be achieved only up to $M_{\rm DM}\sim420$ GeV. Beyond this range of DM mass, the doublet mass becomes $>920$ GeV. A pure doublet fermion DM achieves the correct relic near $M_{\psi}\sim1000$ GeV. In this case, the doublet decay alone produces nearly the required relic while the presence of other $2\rightarrow2$ processes enhances the production leading to an over abundant relic. Therefore, the region right to the contours is excluded as the DM becomes overabundant in this region.  For the same reason, in case $\Delta{M}=1000$ GeV, the maximum DM mass allowed by the relic density constraint is $\sim200$ GeV. Another observation from Fig. \ref{fig:freezein}, is that for DM mass $<1$ TeV with $\Delta{M}>1$ GeV, the correct relic of DM is obtained for $\sin\theta$ varying in the range \{$10^{-13}-10^{-11}$\} and are not excluded by the BBN constraint. For $10^{-7}>\sin\theta>10^{-11}$, the DM is over produced through freeze-in processes.

\subsection{Dark matter production via SuperWIMP mechanism}\label{sec:sumerwimp}
For $\sin\theta<\mathcal{O}(10^{-14})$, the DM relic can be achieved using the SuperWIMP mechanism \cite{Garny:2018ali,Junius:2019dci,Borah:2021rbx}. In this range of $\sin\theta$, $2\rightarrow2$ productions are negligible and DM gets produced from the out-of-equilibrium decay of the doublet. The DM relic can be expressed in terms of the doublet relic as
\begin{eqnarray}
    \Omega_{\rm DM}h^2=\left(\frac{M_{\rm DM}}{M_{\Psi}}\right)\Omega_{\Psi}h^2,\label{eq:relicrat}
\end{eqnarray}
\begin{figure}[h]
	\centering    
	\includegraphics[scale=0.38]{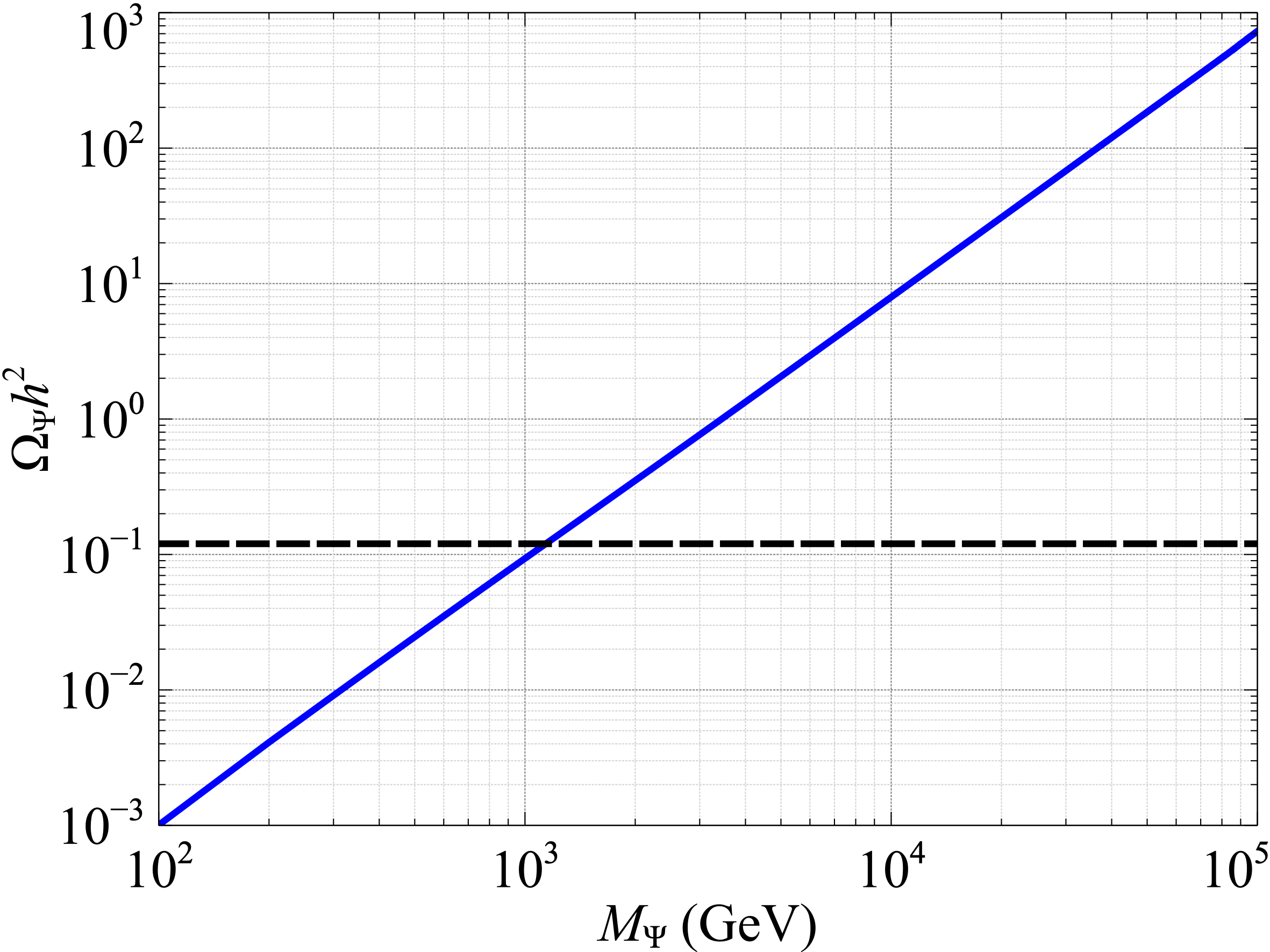}\includegraphics[scale=0.38]{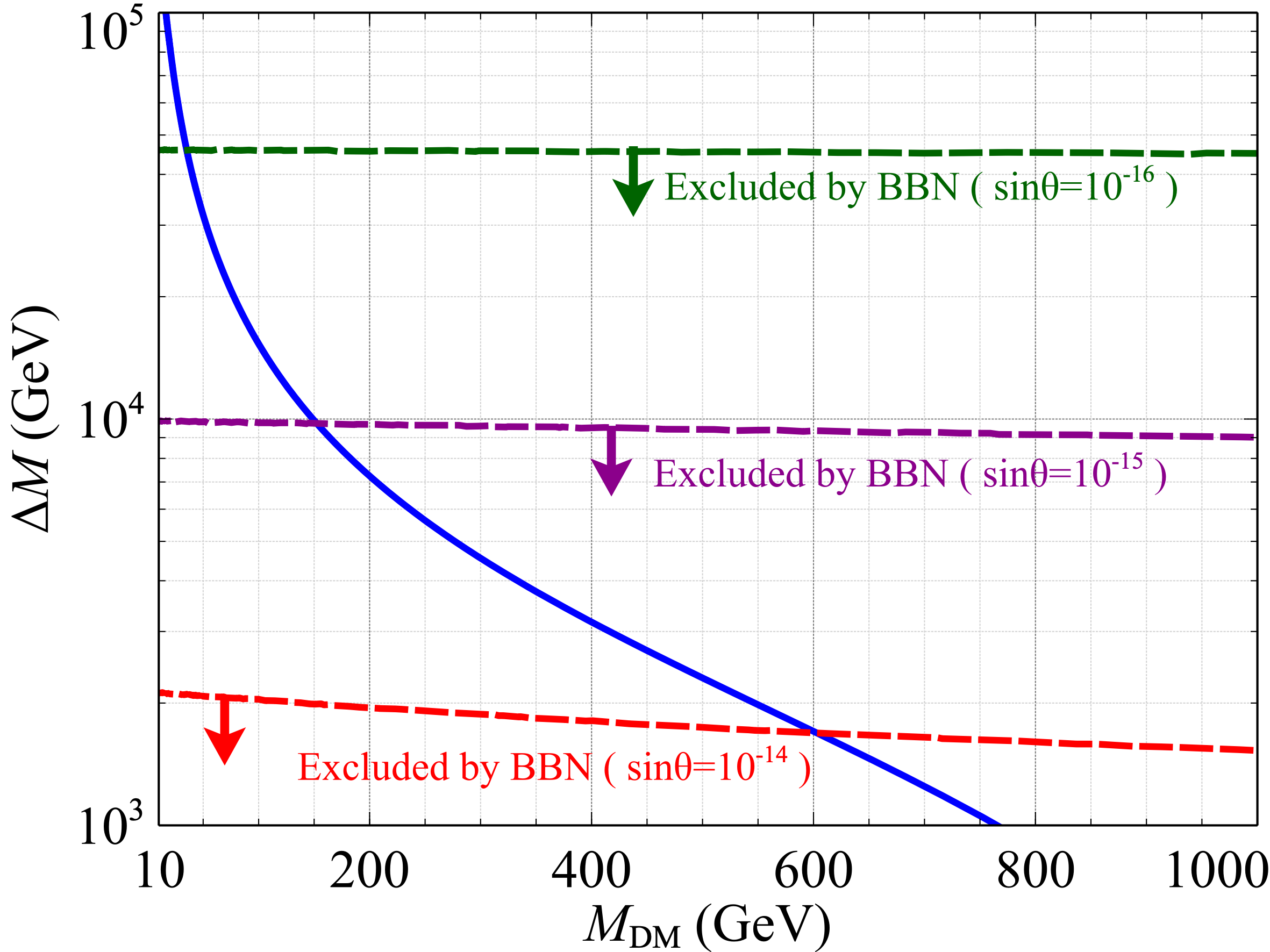}
	\caption{\textit{Left:} Relic abundance of $\Psi$ as function of $M_{\Psi}$. \textit{Right:} Correct relic parameter space obtained via the SuperWIMP mechanism is shown in the plane of $\Delta{M}-M_{\rm DM}$ with the blue line. The colored dashed lines represents the BBN exclusion limits for different $\sin\theta$ as mentioned in the plot.}
	\label{fig:superwimp}
\end{figure}
where $\Omega_{\Psi}h^2$ is the relic of the doublet.\\In the \textit{left} panel of Fig. \ref{fig:superwimp}, we show the relic of the doublet as a function of $M_{\Psi}$. In the \textit{right} panel of Fig. \ref{fig:superwimp}, correct DM relic parameter space is shown in the plane of $\Delta{M}-M_{\rm DM}$ where the $\Delta{M}$ is varied in the range \{$10^3-10^5$\} GeV. Utilizing Eq. (\ref{eq:relicrat}), we calculated the values of $M_{\rm DM}$ that satisfies the correct relic density of the DM. We have also shown the BBN constraints on the decay of the doublet in the plane of $\Delta{M}-M_{\rm DM}$ for 3 different choices of $\sin\theta$ as $10^{-14}$ (red dashed), $10^{-15}$ (dark magenta dashed) and $10^{-16}$ (green dashed). The region below each dashed lines are excluded from the BBN for the corresponding $\sin\theta$.

	\section{Direct detection of dark matter}\label{sec:DD}
	\begin{figure}[h]
		\centering    		\includegraphics[scale=1]{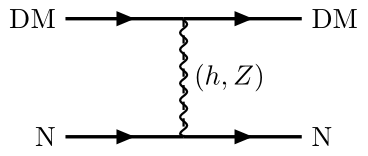}
		\caption{Feynman diagram for spin-independent DM-nucleon scattering cross-section.}
		\label{fig:ddfeyn}
	\end{figure}
	The DM can scatter off the target nucleus via Higgs and Z boson mediated processes as shown in Fig. \ref{fig:ddfeyn}. This leads to the spin-independent (SI) cross-section for Higgs mediated dark matter-nucleon scattering to be
	\begin{equation}
		\begin{aligned}
			\sigma_h^{\rm SI} &= \frac{4}{\pi A^2}\mu^2_r\frac{\Delta M^2\sin^4 2\theta}{2 v^2M^4_h}\Big[\frac{m_p}{v}\Big(f^{p}_{Tu} + f^{p}_{Td} + f^{p}_{Ts} + \frac{2}{9}f^{p}_{TG}+\frac{m_n}{v}\Big(f^{n}_{Tu} + f^{n}_{Td} + f^{n}_{Ts} + \frac{2}{9}f^{n}_{TG}\Big)\Big]^2.
		\end{aligned}
		\label{DDH}
	\end{equation}
	Different coupling strengths between DM and light quarks are
	given by \cite{Bertone:2004pz,Alarcon:2012nr} as $f^p_{Tu}=0.020 \pm 0.004,~f^p_{Td}=0.026 \pm
	0.005,~f^p_{Ts}=0.014 \pm 0.062,~ f^n_{Tu}=0.020 \pm 0.004,~ f^n_{Td}=0.036 \pm 0.005, ~f^n_{Ts}=0.118 \pm 0.062$. The coupling of DM with the gluons
	in target nuclei are parameterized by \cite{Hoferichter:2017olk} $f^{p,n}_{TG}=1-\sum_{q=u,d,s}f^{p,n}_{Tq}$.
	Similarly, the SI DM-nucleon scattering cross-section through Z mediation is given by \cite{ Goodman:1984dc,Essig:2007az}:
	\begin{equation}\label{DDZ}
		\sigma^{\rm SI}_Z = \frac{G^2_F \sin^4\theta}{\pi A^2 }\mu_r^2 \Big|\left[ Z f_p + (A-Z)f_n \right]^2\Big|^2  ,
	\end{equation}
	where the $f_p=f_n=0.33$ corresponds to the form factors for proton and neutron, respectively. Here $\mu_r$ is the reduced mass of the DM-nucleon system, and $A$ and $Z$ are the mass number and atomic number, respectively.
	
\begin{figure}[H]
		\centering    
		\includegraphics[scale=0.5]{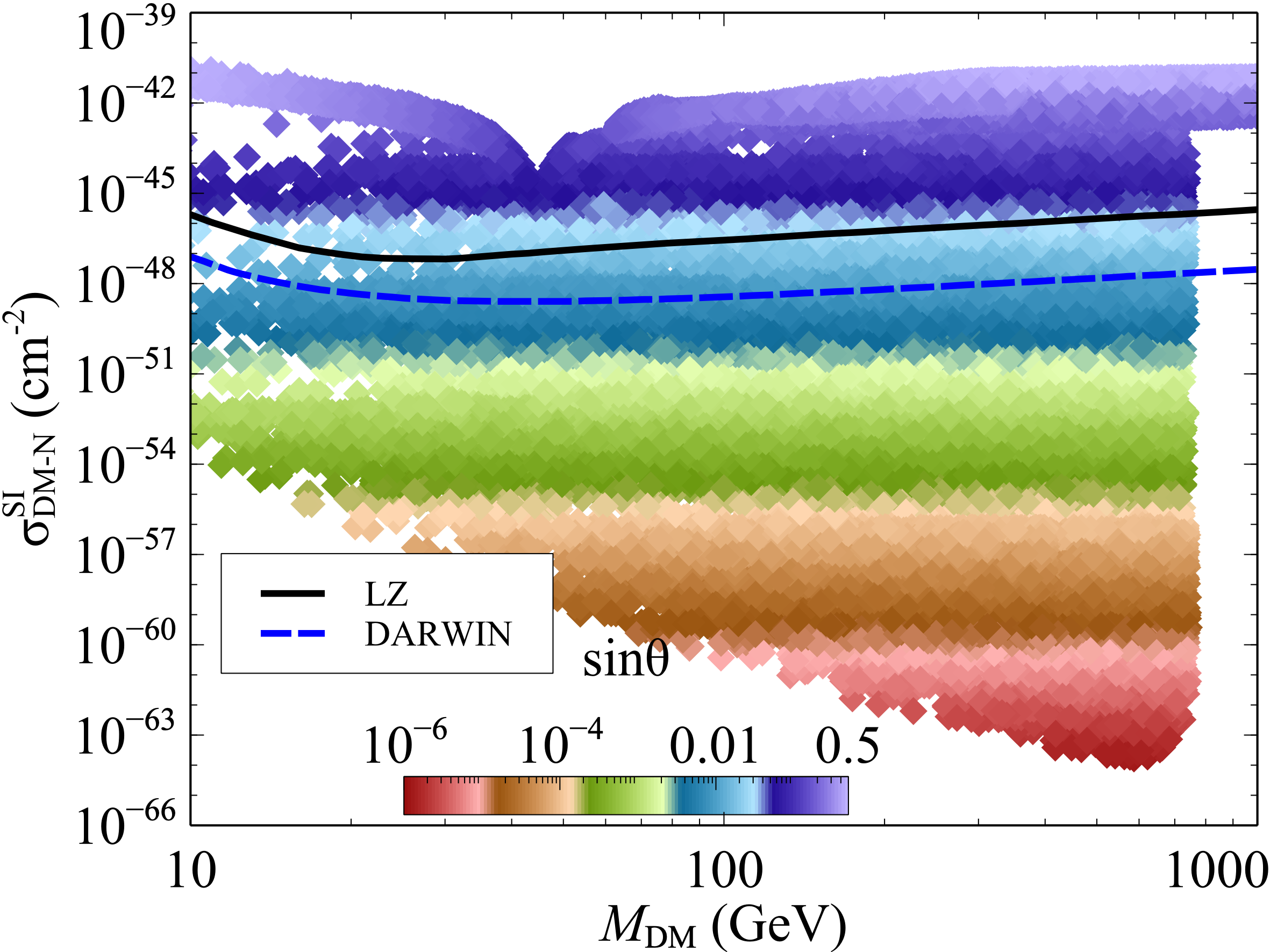}
		\caption{Spin-independent DM-nucleon scattering cross-section as a function of DM mass.}
		\label{fig:dd}
	\end{figure}

	A larger mixing angle leads to an increased direct detection cross-section, which is excluded by the direct detection experiment like LZ \cite{LZ:2022lsv}. We calculated the SI direct detection cross-section ($\sigma^{\rm SI}_{\rm DM-N}=\sigma_h^{\rm SI} +\sigma^{\rm SI}_Z $) for the correct relic points and shown w.r.t $M_{\rm DM}$ in Fig. \ref{fig:dd}. We have also shown the LZ exclusion region in the parameter space of $\sin\theta$ vs $M_{\rm DM}$ in the same plot. 
	It excludes the SD mixing angle for $\sin\theta>0.07$.  The DARWIN \cite{DARWIN:2016hyl} sensitivity is shown with the blue dashed line. When we include the LEP bound, the exclusion goes to $\sin\theta>0.04$.
	
	\section{Displaced vertex signatures}\label{sec:dispacedvertex}
    	\begin{figure}[h]
		\centering    
		\includegraphics[scale=0.5]{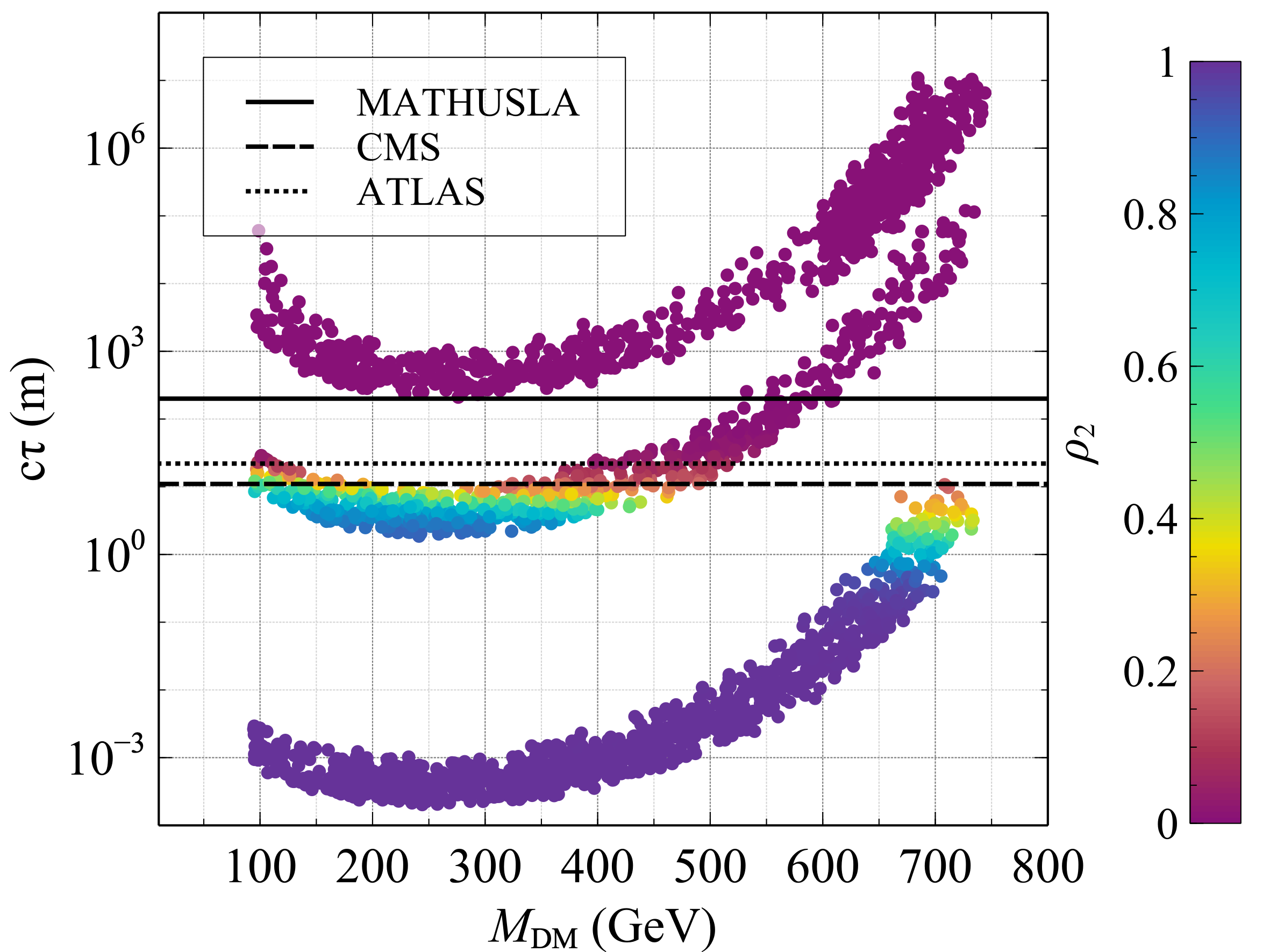}
		\caption{Decay length of the doublet fermion as a function of DM mass with $\rho_2$ in the color code. The mixing angles are varied in the ranges:  $10^{-2}\leq\sin\theta\leq2\times10^{-2}$ (bottom), $10^{-4}\leq\sin\theta\leq2\times10^{-4}$ (middle), and $10^{-5}\leq\sin\theta\leq2\times10^{-5}$ (top). }
		\label{fig:ctau}
	\end{figure}
	Because of gauge interactions, the doublet components, $\psi^\pm$, and $\chi_0$ can be copiously produced in the colliders. Once these particles are produced, they will decay to the DM and charged lepton via the off-shell gauge bosons in the small mass splitting range after traveling some finite distance. The decay length is $\propto(\sin^2\theta\Delta{M}^5)^{-1}$. Thus, one can measure the displaced vertex signatures in present and future collider experiments such as LHC and MATHUSLA. The inclusion of co-scattering can enhance the reach of these experiments in searching for these dark sector particles.
	We have calculated the decay lengths of such particles considering mass splitting ($\Delta M$) in the range from 1 to 12 GeV and shown in Fig. \ref{fig:ctau} with varying DM mass. We considered three ranges of mixing angles: $10^{-2}\leq\sin\theta\leq2\times10^{-2}$ (top), $10^{-4}\leq\sin\theta\leq2\times10^{-4}$ (middle), and $10^{-5}\leq\sin\theta\leq2\times10^{-5}$ (bottom). The color code represents the parameter $\rho_2$. The figure depicts that the inclusion of the co-scattering enhances the lifetime of the dark partners and hence reaches within the sensitivities of LHC and MATHUSLA.
		\begin{figure}[h]
		\centering    
		\includegraphics[scale=0.5]{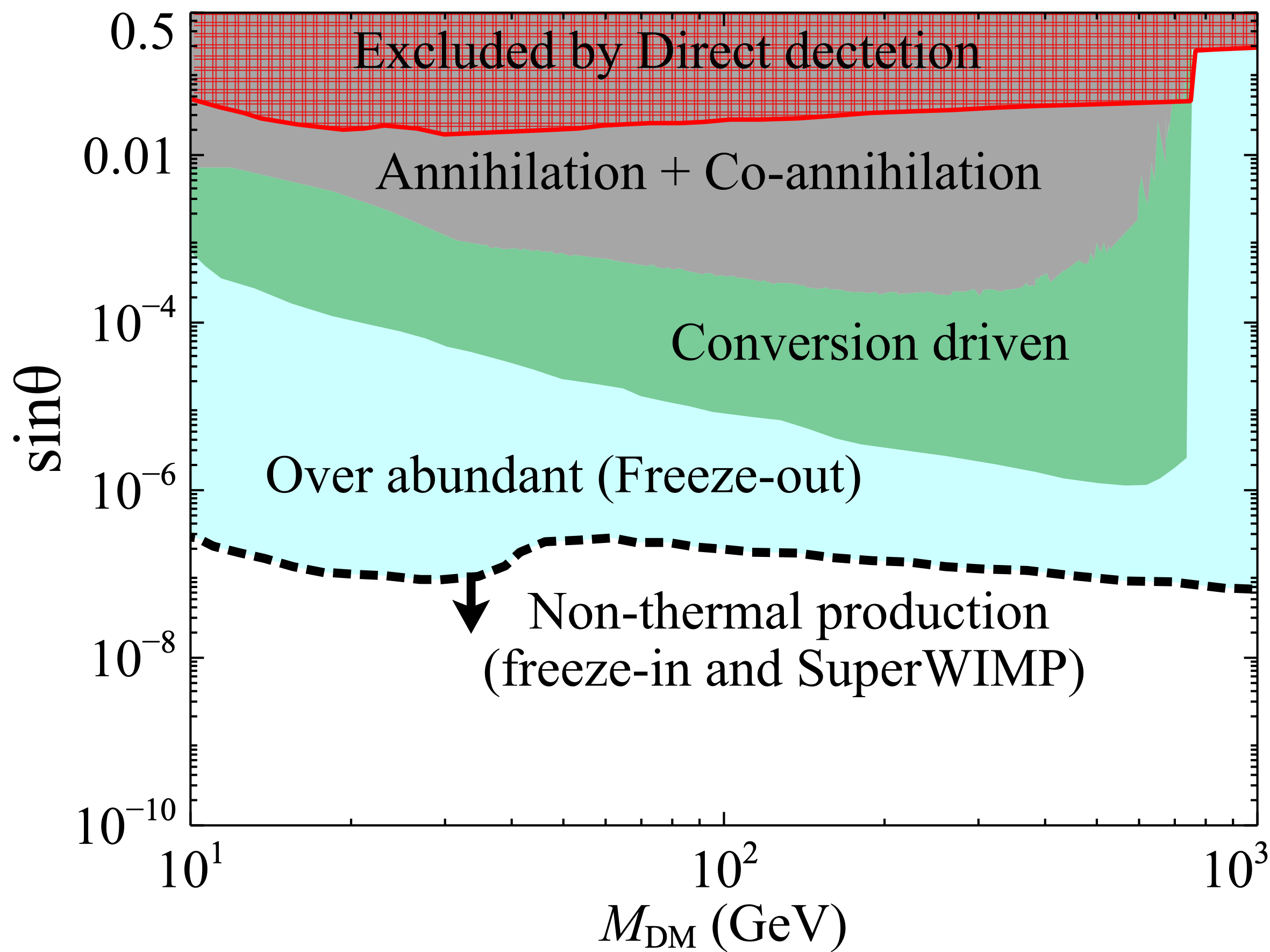}
		\caption{Phase diagram of the SDDM parameter space. The black dashed line separates the region of thermal (upper) and non-thermal (lower) regimes.}
		\label{fig:summary}
	\end{figure}
\section{Conclusions}\label{sec:conclu}
In this work, we explored the full parameter space of singlet-doublet vector-like fermionic dark matter for the first time, accounting for annihilation, co-annihilation, co-scattering, and decay-inverse decay processes for both freeze-out and freeze-in production of DM along with the SuperWIMP mechanism.
	In Fig. \ref{fig:summary}, we show the parameter space in the plane of mixing angle versus DM mass, where the black dashed line separates the thermal and non-thermal regimes at $\Delta M$ = 1 GeV. For $\Delta M<$  1 GeV, this line shifts downward, and for $\Delta M>$  1 GeV, this line moves upward. In the region above the black dashed line, the DM can attain thermal equilibrium, and the relic can be obtained via freeze-out mechanism. On the other hand, below this black dashed line, the DM can never reaches equilibrium, and the relic can be obtained via the non-thermal processes only. The gray-shaded region represents the parameter space where the annihilation and co-annihilation processes solely decide the relic. When the mixing angle is reduced further, the co-scattering processes become the most dominant one, and they decide the DM relic in this regime. This is shown with the green-shaded region.
	As $\sin\theta$ decreases further, the DM decouples from equilibrium earlier, freezing out with a larger abundance. This over-abundant region is depicted as the cyan-shaded area. When the mixing angle drops below $<2\times10^{-7}$, the DM never reaches equilibrium. Below the black dashed line, $\sin\theta$ up to $\mathcal{O}(10^{-11})$, DM relic gets over populated by the freeze-in channels. In the range, $\mathcal{O}(10^{-11})\gtrsim\sin\theta\gtrsim\mathcal{O}(10^{-13})$, the DM relic can be achieved via the freeze-in mechanism. For further smaller $\sin\theta$, typically $<\mathcal{O}(10^{-14})$, the DM relic can be obtained through the SuperWIMP mechanism.
	
	Direct detection experiments constrain the SD mixing angle to values below 0.07, and the LEP bounds on the doublet fermion further tighten this exclusion to 0.04. We have also investigated the displaced vertex signatures of our model at the LHC and MATHUSLA. Our analysis shows that including co-scattering processes significantly expands the parameter space accessible to both LHC and MATHUSLA experiments.

	\appendix
	
	\section{Evolution of abundances for point $A$ in Table \ref{tab:tab1}}\label{app:evo_plot}

\begin{figure}[H]
		\centering
		\includegraphics[scale=0.38]{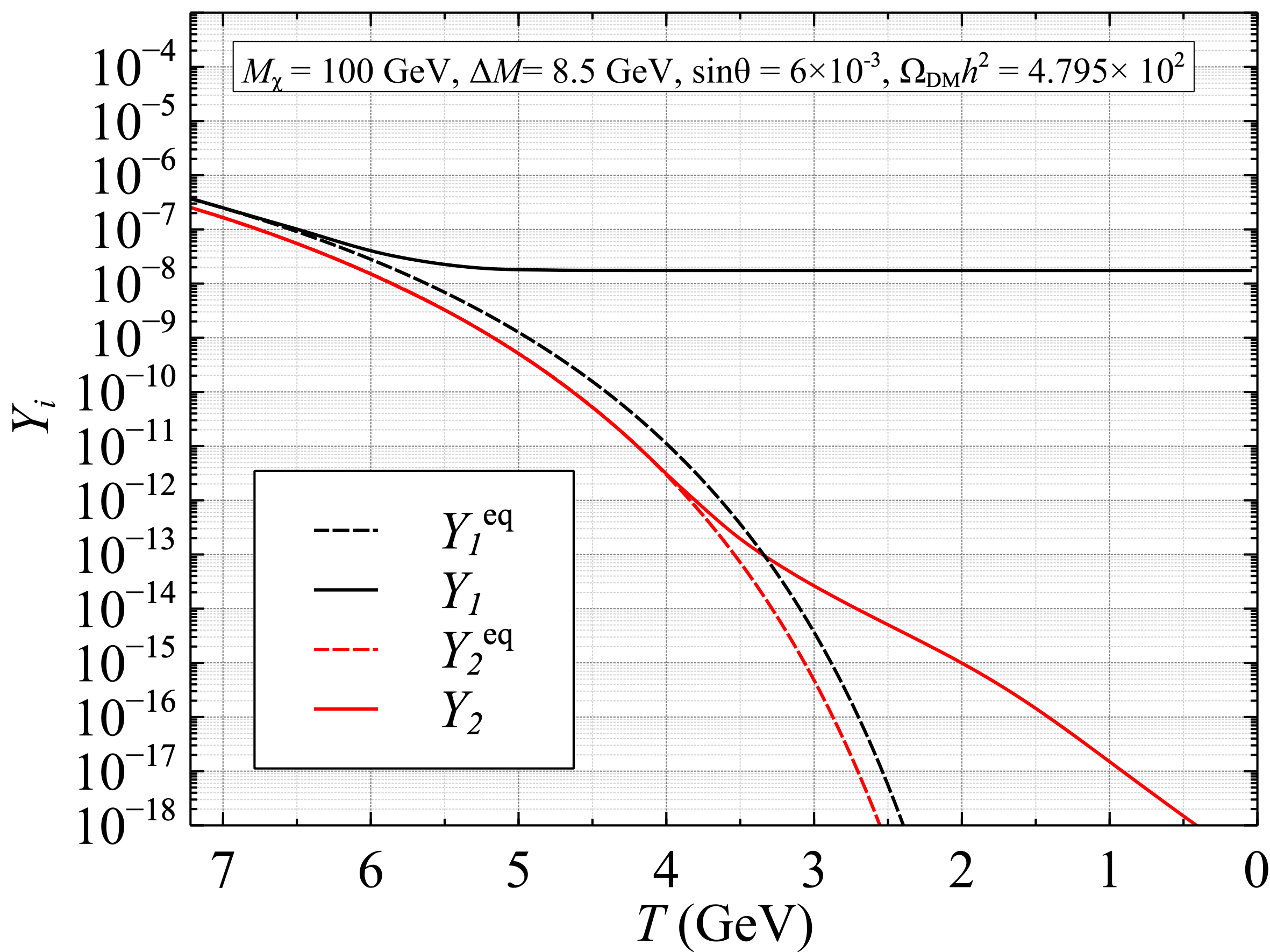}
		\includegraphics[scale=0.38]{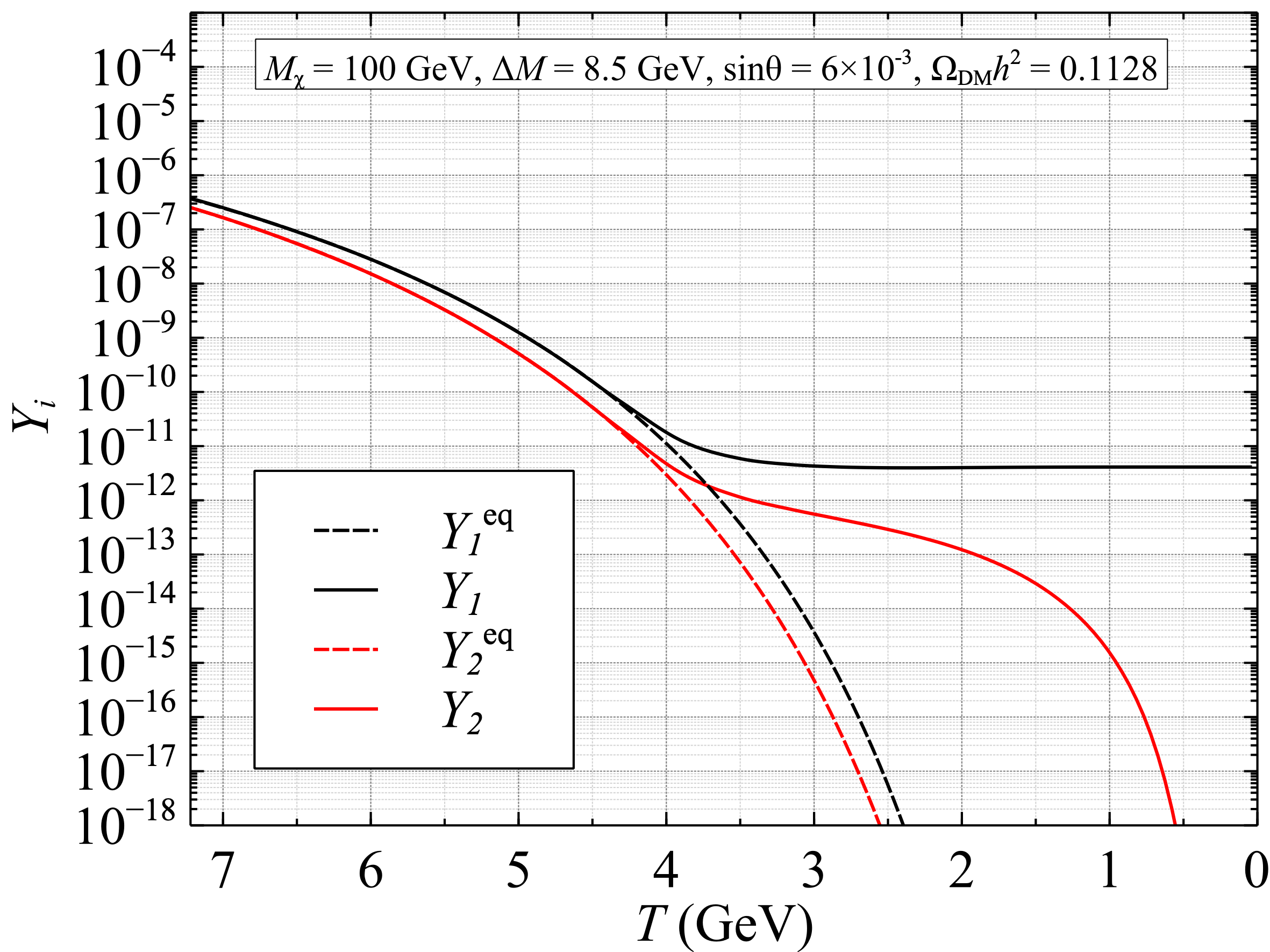}    \caption{\textit{Left:} Evolution of abundances for the point $A$ in Table \ref{tab:tab1} without considering decay. As the $\sin\theta$ is small, once the co-annihilation processes decouple, the singlet decouples early with a larger abundance while the doublet remains in equilibrium for a longer epoch resulting in a smaller abundance. \textit{Right:} Evolution of abundances for the point $A$ in Table \ref{tab:tab1} including decay. Here in presence of decay and inverse decay, the equilibration is maintained for longer duration. This results in correct relic density of the DM.}\label{fig:ev_point_A}
	\end{figure}

    Here we would like to analyze the dependency of the relic for point $A$ in Table \ref{tab:tab1} on the decay and inverse decay terms as given in Eq. (\ref{eq:gamma21}) by switching off the co-scattering term. In Fig. \ref{fig:ev_point_A}, we have shown $Y_1,Y_2$ as a function of temperature. The parameters are mentioned in the figure inset. 

\section{Feynman Diagrams of the processes involved in relic density}

\subsection{Annihilation of dark matter}\label{app:ann}
\begin{figure}[H]
		\centering    
		\includegraphics[scale=0.8]{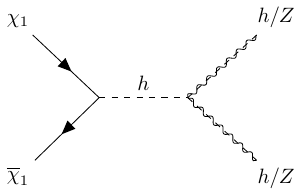}
		\includegraphics[scale=0.8]{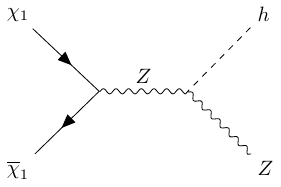}
		\includegraphics[scale=0.8]{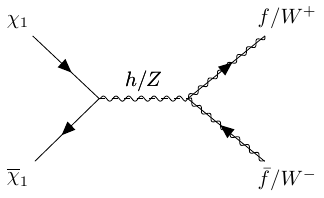}
		\includegraphics[scale=0.9]{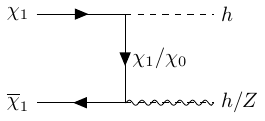}
        \includegraphics[scale=0.9]{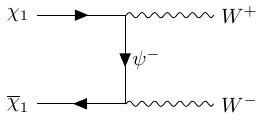}
		\includegraphics[scale=0.9]{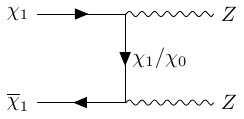}
        
		\caption{DM annihilating to the SM particles through 1100 processes.}
		\label{fig:anndiag}
	\end{figure}

The sector 1 particle (\textit{i.e.} the DM) annihilate to the SM particles and the relevant processes are listed in Fig. \ref{fig:anndiag}.

\subsection{Annihilation and co-annihilation among the sector 2 particles}\label{app:ann_doublet}
\begin{figure}[H]
		\centering    
		\includegraphics[scale=0.8]{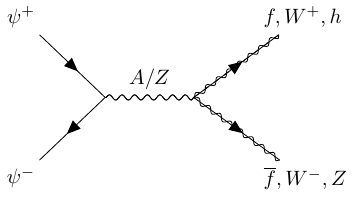}
		\includegraphics[scale=0.8]{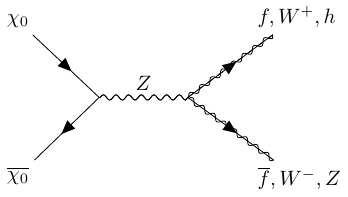}
        \includegraphics[scale=0.8]{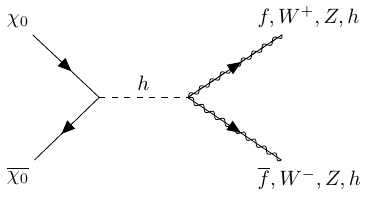}
        \includegraphics[scale=0.9]{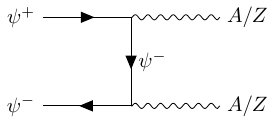}
		\includegraphics[scale=0.9]{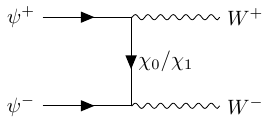}\\
		\includegraphics[scale=0.9]{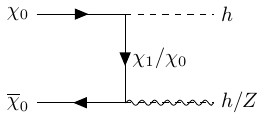}
        \includegraphics[scale=0.9]{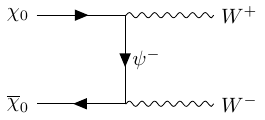}
		\includegraphics[scale=0.9]{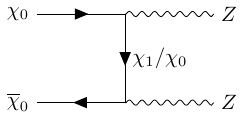}
        \includegraphics[scale=0.9]{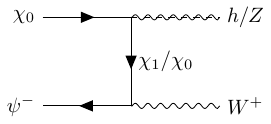}
		\includegraphics[scale=0.9]{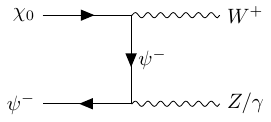}
        \includegraphics[scale=0.7]{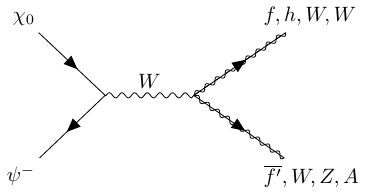}
        
		\caption{Annihilation and co-annihilation of the doublet components to the SM particles through 2200 processes.}
		\label{fig:anndoubletdiag}
	\end{figure}

The sector 2 particles (\textit{i.e.} the doublet components) annihilate and co-annihilate to the SM particles and the relevant processes are listed in Fig. \ref{fig:anndoubletdiag}.

\subsection{Co-annihilation processes}\label{app:coann}
\begin{figure}[H]
		\centering    
		\includegraphics[scale=0.8]{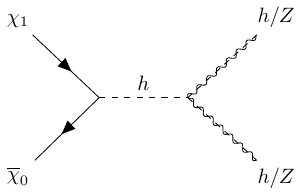}
		\includegraphics[scale=0.8]{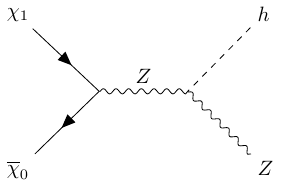}
		\includegraphics[scale=0.8]{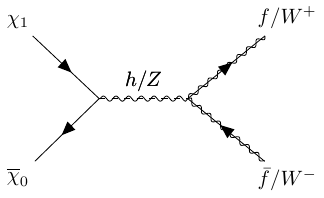}
		\includegraphics[scale=0.9]{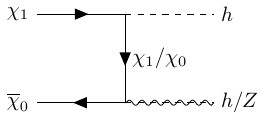}
        \includegraphics[scale=0.9]{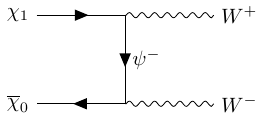}
		\includegraphics[scale=0.9]{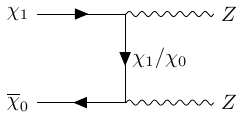}
        
		\caption{DM co-annihilating with the neutral component of the doublet through 1200 processes.}
		\label{fig:coanndiag1}
	\end{figure}

\begin{figure}[H]
		\centering    
		\includegraphics[scale=0.8]{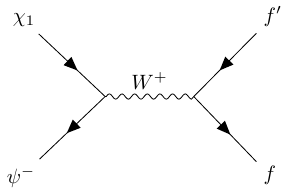}
		\includegraphics[scale=0.8]{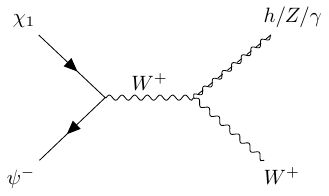}\\
		\includegraphics[scale=0.9]{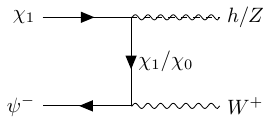}
		\includegraphics[scale=0.9]{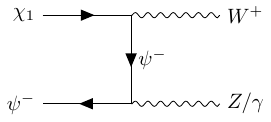}
        
		\caption{DM co-annihilating with the charged component of the doublet through 1200 processes.}
		\label{fig:coanndiag2}
	\end{figure}

The sector 1 particle co-annihilate with the sector 2 particles to the SM particles. The relevant processes are listed in Fig. \ref{fig:coanndiag1} and Fig. \ref{fig:coanndiag2}.
	
\subsection{Co-scattering processes}\label{app:co-scattering}

\begin{figure}[h]
		\centering    
        \includegraphics[scale=0.8]{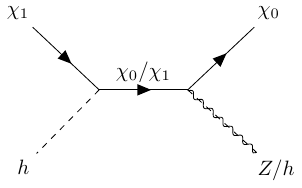}
    \includegraphics[scale=0.8]{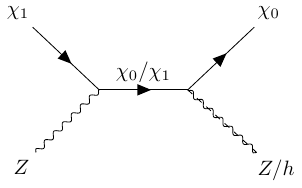}
  \includegraphics[scale=0.8]{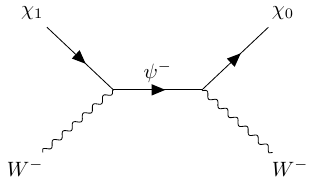}
		    \includegraphics[scale=0.9]{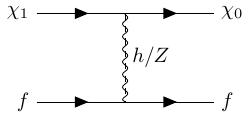}
		\includegraphics[scale=0.9]{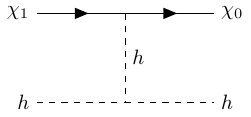}
        \includegraphics[scale=0.9]{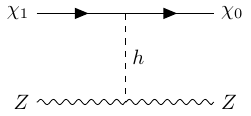}
		\includegraphics[scale=0.9]{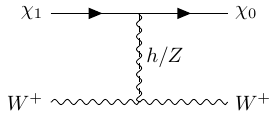}\\
		\includegraphics[scale=0.9]{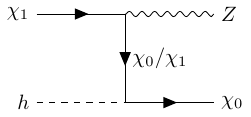}
		\includegraphics[scale=0.9]{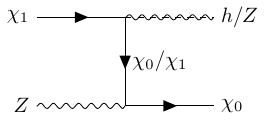}
		\includegraphics[scale=0.9]{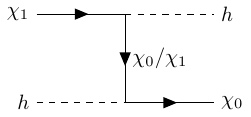}
		\includegraphics[scale=0.8]{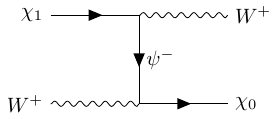}
        \includegraphics[scale=0.8]{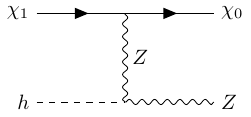}
        \includegraphics[scale=0.8]{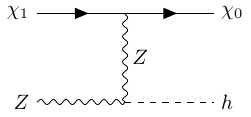}
        
		\caption{DM co-scattering with the SM bath particles to the neutral component of the doublet state through 1020 processes.}
		\label{fig:coscadiag1}
	\end{figure}

\begin{figure}[H]
		\centering    
		\includegraphics[scale=0.8]{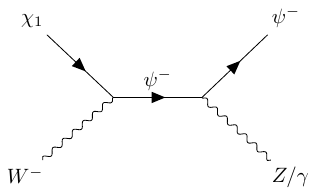}
        \includegraphics[scale=0.8]{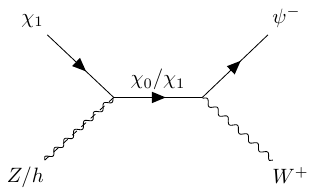}
        
    \includegraphics[scale=0.8]{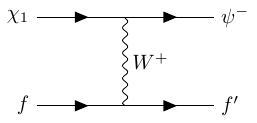}
		\includegraphics[scale=0.8]{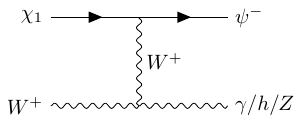}
		\includegraphics[scale=0.8]{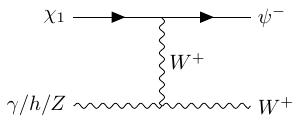}
        \includegraphics[scale=0.8]{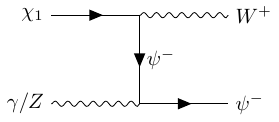}
		\includegraphics[scale=0.8]{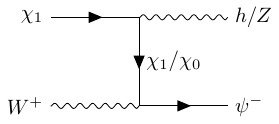}
    
		\caption{DM co-scattering with the SM bath particles to the charged component of the doublet state through 1020 processes.}
		\label{fig:coscadiag2}
	\end{figure}

The sector 1 particle scatters with the SM particles to the sector 2 particles and SM particles known as the co-scattering processes. The relevant processes are listed in Fig. \ref{fig:coscadiag1}, and Fig. \ref{fig:coscadiag2}.

	\section{Decay rates}\label{app:decayrate}

\begin{figure}[h]
		\centering    
		\includegraphics[scale=0.05]{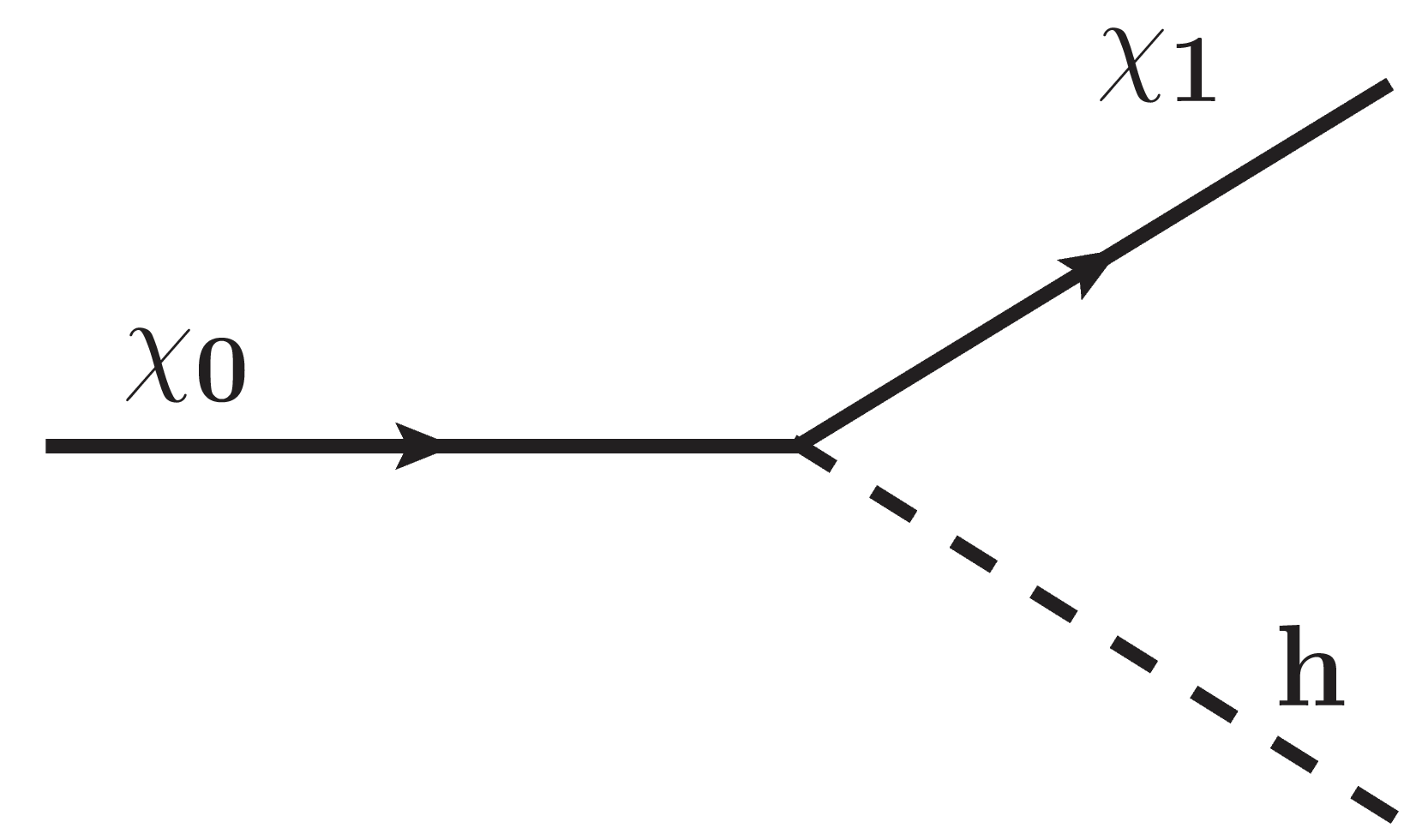}
		\includegraphics[scale=0.05]{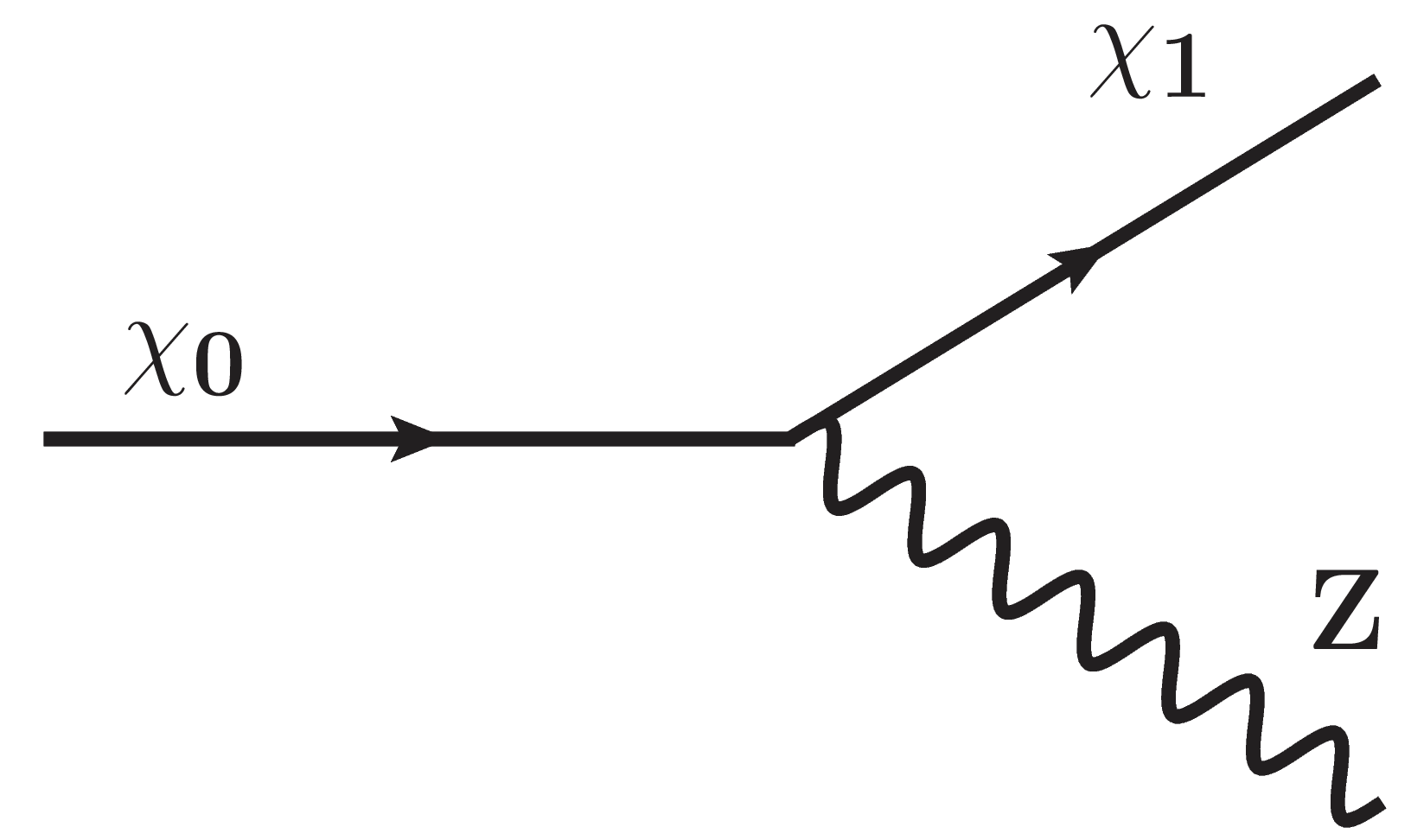}
		\includegraphics[scale=0.05]{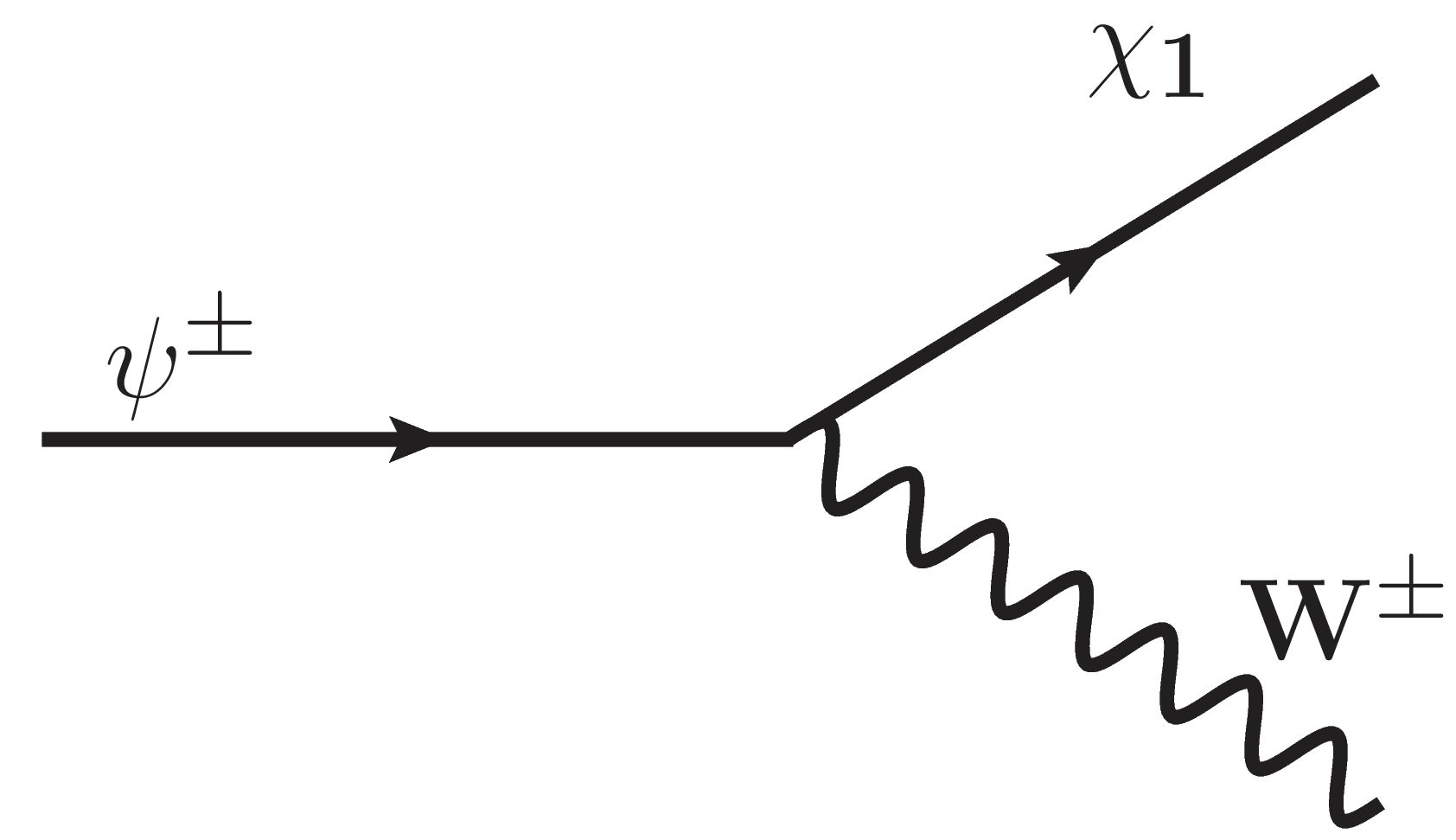}
		\includegraphics[scale=0.25]{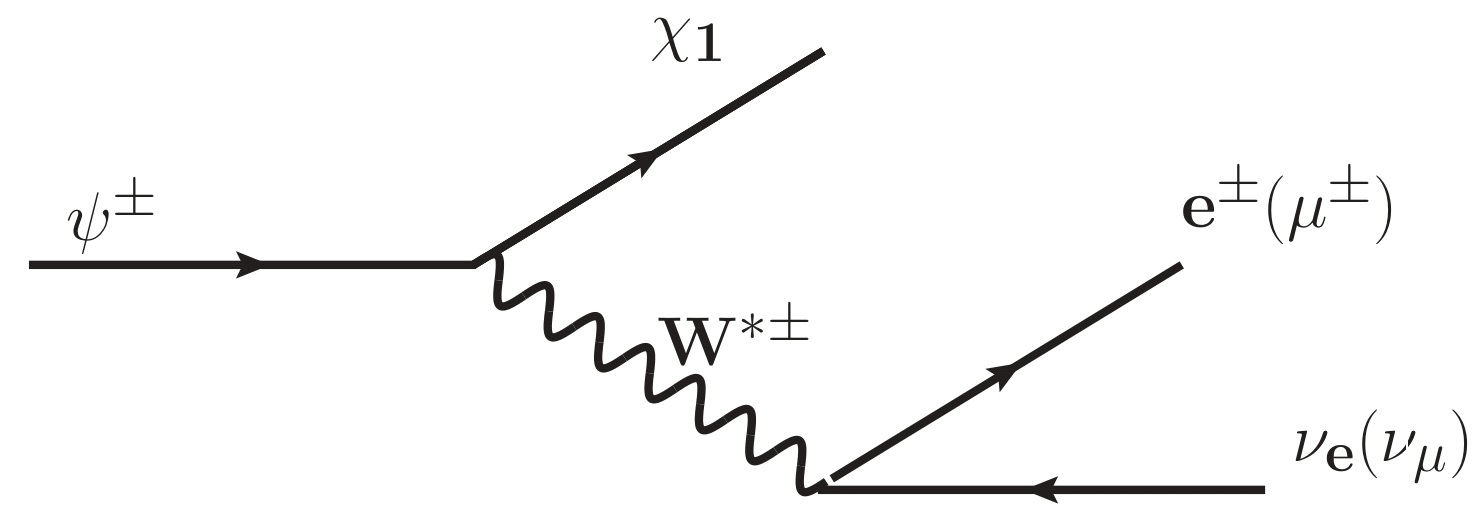}
		\caption{The Feynman diagrams for the production of the DM from the decay of sector 2 particles.}
		\label{fig:dmdecay}
	\end{figure}

The kinematically allowed two and three body decay modes of the charged and neutral component of the doublet are shown in Fig. \ref{fig:dmdecay}.
    
	Two body decay rates are given as
	
	\begin{eqnarray}
		\Gamma_{a\rightarrow bc}&=&\frac{\sqrt{\Lambda(M_a^2,M_b^2,M_c^2)}}{2 M_a}\frac{1}{32 \pi^2M_a^2}4\pi\left|\mathcal{M}_{a\rightarrow bc}\right|^2,\\
		{\rm where,~}\Lambda(x,y,z)&=&x^2+y^2+z^2-2xy-2yz-2xz
	\end{eqnarray}
	
	\begin{eqnarray}
		\left|\mathcal{M}_{\psi^0\rightarrow\chi H}\right|^2&=&\frac{(M_{\chi_0}-M_{\chi_1})^2\sin^22\theta}{4v^2}\left(1-2\sin^2\theta\right)^2\left(M_{\chi_1}^2+2M_{\chi_0}M_{\chi_1}+M_{\chi_0}^2-M_h^2\right),\\
		\left|\mathcal{M}_{\psi^0\rightarrow\chi Z}\right|^2&=&G_F\left(1-\sin^2\theta\right)\sin^2\theta(M^4_{\chi_1}-2M^2_{\chi_0}M^2_{\chi_1}+M^2_{\chi_1}M^2_Z-6M_{\chi_0}M_{\chi_1}M^2_Z+M^4_{\chi_0}\nonumber\\&&+M^2_{\chi_0}M^2_Z-2M^4_Z),\\
		\left|\mathcal{M}_{\psi^-\rightarrow\chi W}\right|^2&=&\frac{G_F}{2}\sin^2\theta(M^4_{\chi_1}-2M^2_{\psi^-}M^2_{\chi_1}+M^2_{\chi_1}M^2_W-6M_{\psi^-}M_{\chi_1}M^2_W+M^4_{\psi^-}\nonumber\\&&+M^2_{\psi^-}M^2_Z-2M^4_Z).
	\end{eqnarray}
	
	The three body decay rates are \cite{Cirelli:2005uq}
	
	\begin{eqnarray}
		\Gamma_{\psi\rightarrow\chi l\nu_l}=\frac{2G_F^2}{15\pi^3}(\Delta M )^5
		\label{eq:3bodydecay}
	\end{eqnarray}

	\acknowledgments{P.K.P. would like to acknowledge the Ministry of Education, Government of India, for providing financial support for his research via the Prime Minister’s Research Fellowship (PMRF) scheme. The work of N.S. is supported by the Department of Atomic Energy-Board of Research in Nuclear
	Sciences, Government of India (Ref. Number: 58/14/15/2021- BRNS/37220). We would like to thank Alexander Pukhov for his assistance on $\tt micrOMEGAs$. We would like acknowledge the anonymous referee of JCAP for his/her constructive comments
	which improved the manuscript.}

\providecommand{\href}[2]{#2}\begingroup\raggedright\endgroup

\end{document}